\documentclass[11pt, onecolumn]{article}
\usepackage[margin = 1in]{geometry}

\usepackage{authblk}
\usepackage{amssymb}
\usepackage{booktabs}
\usepackage{amsmath}

\usepackage{tikz}
\usepackage{pgf}
\usetikzlibrary{patterns,arrows,decorations.pathreplacing}

\usepackage{amsthm}
\usepackage{url}
\usepackage{algorithm}
\usepackage{algorithmic}
\usepackage{trivfloat}
\trivfloat{program}

\usepackage{float}
\usepackage{epstopdf}
\usepackage{verbatim}
\usepackage{empheq}
\usepackage{hyperref}
\usepackage{color}

\usepackage[us, 12hr]{datetime}

\usepackage{multirow}
\usepackage{setspace}
\newcommand{\tp}{\mathsf{T}}							
\newcommand{\ct}{\mathsf{H}}							

\renewcommand{\Re}{\mathsf{Re}}
\renewcommand{\Im}{\mathsf{Im}}

\newcommand{\real} {\mathbb{R}}
\newcommand{\complex} {\mathbb{C}}

\newcommand{\ii} {\mathrm{i}}

\newcommand{\ds}{\displaystyle}

\newcommand{\bnorm }{\theta}			%

\newcommand{\opt}{^{\star}}

\newcommand{\lam} { \lambda}	
\newcommand{\eps} { \epsilon}	
\newcommand{\phie} {\phi_{\epsilon}}

\newcommand{\half}{\frac{1}{2}}					%

\providecommand{\norm}[1]{\lVert#1\rVert}
\providecommand{\abs}[1]{\left\vert #1 \right\vert}

\newtheorem{theorem}{Theorem}

\providecommand{\tabref}[1]{Table~\ref{#1}}

\providecommand{\figref}[1]{Figure~\ref{#1}}
\providecommand{\appref}[1]{Appendix~\ref{#1}}
\providecommand{\theref}[1]{Theorem~\ref{#1}}
\providecommand{\eqnref}[1]{\eqref{#1}}

\newcommand{\Figurescale}		{0.625}				
\newcommand{\figurescale}		{0.475}				
\newcommand{\figurescalesmall}	{0.375}				

\allowdisplaybreaks

\definecolor{mygray}{rgb}{0.725,0.725,0.725}

\title{Detection of Faults in Rotating Machinery Using \\ Periodic Time-Frequency Sparsity%
\footnote{pre-print submitted to Journal of Sound and Vibration}}

\author[1]{Yin Ding\thanks{email: yd372@nyu.edu}}
\author[2]{Wangpeng He}
\author[4]{Binqiang Chen}
\author[3]{Yanyang Zi}
\author[1]{Ivan W. Selesnick}

\affil[1]{Tandon School of Engineering, New York University, 6 Metrotech Center, Brooklyn, NY 11201, USA}
\affil[2]
	{School of Aerospace Science and Technology, Xidian University, Xi'an, China} 
\affil[3]{School of Mechanical Engineering, Xi'an~Jiaotong University, Xi'an, China}
\affil[4]{School of Aeronautics and Astronautics, Xiamen University, Xiamen, China}

\begin{document}
\maketitle
\begin{abstract}

This paper addresses the problem of extracting periodic oscillatory features in vibration signals for detecting faults in rotating machinery.
To extract the feature, we propose an approach in the short-time Fourier transform (STFT) domain
where the periodic oscillatory feature manifests itself as a relatively sparse grid.
To estimate the sparse grid,
we formulate an optimization problem using customized binary weights in the regularizer,
where the weights are formulated to promote periodicity.
In order to solve the proposed optimization problem,
we develop an algorithm called augmented Lagrangian majorization-minimization algorithm,
which combines the split augmented Lagrangian shrinkage algorithm (SALSA)  with
majorization-minimization (MM),
and is guaranteed to converge for both convex and non-convex formulation.
As examples,
the proposed approach is applied to simulated data,
and used as a tool for diagnosing faults in bearings and gearboxes for real data,
and compared to some state-of-the-art methods.
The results show the proposed approach can effectively detect and extract the periodical oscillatory features.
\end{abstract}



\section{Introduction}

Condition monitoring and fault diagnosis of rotating machines are of great importance to prevent machinery breakdown in numerous industrial applications.
Vibration signals generated by faults in rotating components have been widely studied.
Detecting and extracting transient vibration signatures is of vital importance for vibration-based detection of faults in rotating machinery.
However, the observed vibration signals are usually corrupted by very heavy background noise \cite{randall2011rolling, feng2012vibration}.
Many diagnostic techniques have been developed to extract the fault features based on different transforms,
such as short time Fourier transform (STFT) \cite{fd_Yang_mssp_2005, feng2013recent,fd_Kar_jsv_2008, fd_He_jsv_2013, fd_Qin_jsv_2013}
and wavelet transform
\cite{yan2014wavelets, chen2012fault, yan2010harmonic, fd_Kar_mssp_2006, fd_Zhen_jsv_2008, he2013vibration, he2014automatic, fd_Bozchalooi_jsv_2007,
 fd_Kankar_2011}.
Some methods use morphological decomposition methods,
such as empirical mode decomposition (EMD) \cite{lei2013review, guo2013novel},
or demodulation \cite{fd_Sheen_JSV_2004, fd_Liang_mssp_2010, fd_Wang_jsv_2014}.
Some methods combine conventional time-frequency analysis with other techniques suitable for transients detection,
such as bispectrum-based method \cite{fd_Benbouzid_2000, fd_bispectrum_Chow_1995}
and spectral kurtosis (SK) based methods \cite{antoni2006spectral, antoni2007fast, zhang2009rolling, fd_Chen_mssp_2013}.

The use of sparse representations for fault detection was initially illustrated in Ref.~\cite{fd_Yang_mssp_2005},
where basis pursuit denoising (BPD) \cite{Chen_1994_BP, Chen_SIAM_1998} was adopted to exploit the sparsity of vibration signals in different domains in order to detect and extract fault features.
In Ref.~\cite{fd_Qin_jsv_2013}, the author considered a morphological component analysis (MCA) problem \cite{Elad_etal_2005_ACHA, mca_Starck_2004, Starck_2005_TIP},
which is a sparsity-based decomposition approach, to improve the extraction of fault features.
In Ref.~\cite{fd_Cui_jsv_2014} matching pursuit (MP) \cite{Mallat_1993} was used on a customized dictionary to extract oscillatory fault features.
In Ref.~\cite{he2013tunable}, a sparse representation using the tunable Q-factor wavelet transform \cite{Selesnick_TQWT_2011, Selesnick_2011_SPIE_TQWT}
is utilized to extract oscillatory fault features.
Most recently, sparse coding techniques are introduced for fault feature extraction \cite{tang2014sparse}.

The periodicity (also called fault characteristic or fundamental frequency in some cases) of potential fault features
is very important information for fault diagnosis.
In many cases, this information can be simply obtained using the geometry of the specific components under steady rotating speed,
or directly obtained from the user operation manual.
It is already required as necessary information in many fault diagnosis methods \cite{yan2014wavelets, fd_Liang_mssp_2010, fd_Su_mssp_2010, fd_Wang_jsv_2015}.
%
%
Some fault diagnosis methods use periodicity information as a verification step after feature extraction (e.g. \cite{fd_Liang_esa_2014,  fd_Qin_jsv_2013}),
and some methods use it to determine an optimal frequency band when extracting the oscillation frequency of the fault feature
\cite{fd_Liang_mssp_2010, fd_Su_mssp_2010, fd_Wang_jsv_2015}.
%
Only a few very recent works use the temporal periodic structure to extract fault features
\cite{fd_McDonald_mssp_2012, He_mssp_2016}.

The overlapping group sparsity (OGS) approach has been discussed in \cite{Chen_Selesnick_2014_OGS, Chen_Selesnick_2014_GSSD},
where a multi-dimensional group structure has been modeled in various domains
and convex optimization problems have been formulated, to take advantage of group behavior (e.g. coefficients in STFT domain)
to recover signals from noisy observations.
Using OGS to extract fault features was introduced and formulated as an convex optimization problem in \cite{He_mssp_2016},
in which the fault features are assumed to be a periodic sequence of pulses.
In addition, it was shown that the objective function can be convex even if
some specified non-convex penalty functions are used to promote sparsity.
Moreover, when the periodicity information is known, a binary-weighted group structure was used in the regularization,
capturing the period of the potential fault features.

In this work, we consider the problem wherein the fault features are comprised of a sequence of oscillatory transients, and sparsity should be promoted in the time-frequency domain.
The discrete-time vibration observation $y$ is modeled by
\begin{align}\label{eqn:fs_model}
	y = x + w,
\end{align}
where $x$ is an approximately-periodic sequence of oscillatory transients where the period is determined by the characteristic frequency of potential faults,
and $w$ is additive Gaussian noise.
Moreover, we assume that the oscillation frequencies of the transient component $x$ are unknown but relatively sparse,
and the periodicity is over the time domain.

Under the signal model, we seek a solution to the optimization problem
\begin{equation} \label{eq:optimization}
	c\opt =\arg \min_{c} \Big \{ F(c)= \half \norm{ y - Ac }_2^2 + \lam \Phi(c) \Big \},
\end{equation}
for the purpose of extracting the fault feature $x\in \real^{N}$,
where $c$ denotes the coefficients corresponding to an overcomplete transform $A$ of $x$, where $x=Ac$,
and $\lam >$ 0 is a regularization parameter and $\Phi$ is a sparsity-promoting penalty function (regularizer) in the domain of $c$.
This paper aims to extract the oscillatory fault features as periodically structured groups of coefficients in the time-frequency domain.
We set the operator $A$ to be the normalized inverse STFT, and the function $\Phi$ is customized to capture the periodicity in the STFT domain.

To solve a relatively complicated sparsity-based optimization problem,
where an iterative scheme has to be adopted,
methods are usually based on two algorithms.
One is based on majorization-minimization (MM)
\cite{FBDN_2007_TIP, mm_Lange_2000, mm_Hunter_tutorial_2004},
such as the algorithms given in 
\cite{FBDN_2007_TIP, Oliveira_sp_2009, Ding_2015_SP, Selesnick_tara_2014, Selesnick_tv_ncvx_2015, Selesnick_spl_2015},
and further the method of iterated soft-thresholding algorithm (ISTA)
\cite{Beck_2009_SIAM, DDDM_2003} can be considered as a special case of MM asa well (majorizing data fidelity term based on Lipschitz constant).
Another trend is based on alternating direction method of multipliers (ADMM)
\cite{Boyd_2011_admm, opt_Eckstein_1992},
and its extension such as
split augmented Lagrangian shrinkage algorithm (SALSA) \cite{Afonso_2010_TIP_SALSA},
 and some recent applications using such algorithm are presented in \cite{Ning_QRS_2013, Selesnick_2011_SPIE_TQWT}.

Some algorithms can be explained as a combination of ADMM and MM as a special case.
For instance, a specific function of Bregman distance was described in \cite{opt_Osher_2005},
which is similar to a majorizer,
utilized to derive an iterative method for total variation problem \cite{ROF_1992}.
Such concept has been used on total variation problem with multiple extensions
such as combinations with wavelet-based denoising, compressive sensing 
\cite{tv_Xu_tip_2007, cs_Yin_2008, cs_Ma_2008}.
Moreover, the Bregman distance function in \cite{opt_Osher_2005}
was used to in other iterative algorithms for convex problems as well,
such as the methods proposed in \cite{opt_Zhang_2010, opt_Zhang_jsc_2010}.
Some recent work in the field of optimization also consider this problem.
In \cite{opt_Li_admm_preprint}, a framework is given that instead of 
using the Bregman distance function in \cite{opt_Osher_2005}, but using proper `indefinite proximal terms' to majorize augmented Lagrangian,
the resulting majorized ADMM-style algorithm will converge for convex problem.
Moreover, in \cite{opt_Chen_admm_2015_preprint}, 
an ADMM-style algorithm with a embedded criteria checking step is proposed based on majorizing the corresponding
augmented Lagrangian as well.
Additionally, in \cite{opt_Cui_admm_preprint}, a scheme of majoring a cross term with the ADMM-style iteration is proposed for convex problem.

In this work,
we propose an algorithm that combines ADMM and MM with a more general and simpler formulation,
based on block successive upper-bound minimization (BSUM) algorithm \cite{opt_Razaviyayn_2013}.
Moreover,
the algorithm allows non-convex regularization to promote the resulting sparsity, 
and it is guaranteed to converge to a local minimum even though the objective function is not convex.

\subsection{Notation}

In this paper, the elements of a vector $ x \in \real^N$ are denoted as $x_n$ or $[x]_n$,
and the norms are defined as
\begin{subequations}
\begin{align}
	\norm{x}_1 		& : = \sum_{n} \abs{x_n}		, \\
	\norm{x}_2^2 	& : = \sum_{n} \abs{x_n}^2	.
\end{align}
\end{subequations}
The norms of a complex vector $ z \in \complex^{N}$  is defined as
\begin{subequations}
\begin{align}
	\norm{z}_1 		& : = \sum_{n} \abs{z_n}		, \\
	\norm{z}_2^2 	& : = \sum_{n} \abs{z_n}^2	,
\end{align}
\end{subequations}
where the definition of absolute value of a complex number $a \in \complex$ is given by
\begin{align}
	\abs{a} 
	: = \sqrt{ [\Re(a)]^2 + [\Im(a)]^2  },
\end{align}
where $\Re(\cdot)$ and $\Im(\cdot)$ are operators taking the real and imaginary part of a complex number respectively.
As a consequence, $\norm{z}_2^2 $ can be written explicitly as
\begin{subequations}
\begin{align}
	\norm{z}_2^2
		&  = \sum_n  [\Re([z]_n)]^2 + [\Im([z]_n)]^2 \\
		& = \norm{z_r}_2^2 + \norm{z_i}_2^2,		
\end{align}	
\end{subequations}
where $z = z_r + \ii z_i$, and $ z_r, z_i \in \real^{N} $ are vectors of the real and imaginary part of $z$, respectively.

For multi-dimensional matrices, e.g. $B \in \real^{M \times N}$, we denote its elements as $[B]_{m,n}$.
We use bold $\mathbf{0}$ and $\mathbf{1}$ to express all $0$'s and $1$'s matrices.
More specifically, $\mathbf{0}(K,N)$ is a matrix of size $K \times N$, and all its elements are zero,
and $\mathbf{1}(K,N)$ is a matrix, the elements of which are all one.

\subsection{Review of Majorization-minimization}

The majorization-minimization (MM) method \cite{FBDN_2007_TIP, mm_Lange_2000, mm_Hunter_tutorial_2004}
simplifies a complicated optimization problem into a sequence of easier ones.
To solve a convex optimization problem
\begin{align}
	u \opt = \arg \min_{u} F(u),
\end{align}
the MM method is described by iteration
\begin{align}\label{eqn:bsum_mm_iteration}
	u^{(i+1)} = \arg \min_{u} G( u , u^{(i)} ),
\end{align}
where $G : \real^{N} \times \real^{N} \to \real$ is a continuously differentiable function, satisfying
\begin{subequations}\label{eqn:bsum_mm_condition}
\begin{align}
	&G(u,v) \ge F(u),\quad \text{ for all } u, v,\\
	&G(u,u) = F(u).
\end{align}
\end{subequations}
The detailed proof of convergence for convex problems has been given in Ref.~\cite[Chapter~10]{mm_Lange_2000}.

\subsection{Review of BSUM}

Block successive upper-bound minimization algorithm (BSUM) is described in \cite{opt_Razaviyayn_2013}.
Its original goal is to overcome the uniqueness requirement of block coordinate descent (BCD) method  (also known as Gauss-Seidel method) \cite{opt_Beck_BCD_2013, opt_Luo_BCD_1992}.
BSUM is an iterative method to solve optimization problems,
which allows the objective function to be non-convex and/or non-smooth.
When the objective function is neither convex nor smooth, the algorithm will at least converge to a stationary point under some weak assumptions \cite[Theorem 2]{opt_Razaviyayn_2013}.
This algorithm has been applied to transceiver design
in MIMO multi-cellular communication system, which has a non-convex formulation \cite{Razaviyayn_TSP_2014}.
In order to facilitate our proposed algorithm, as a short review, we only rewrite a special case of BSUM using our notation.

The problem is defined as
\begin{align}\label{eqn:bsum_bsum_cost}
	x\opt = \arg \min_{x} Q(x) ,
\end{align}
where $Q : \real^{N} \to \real$ is continuous and has at least one stationary point.
In this case, we slightly relax the definition of `optimal solution',
whereas if the objective function is not convex, we allow the `optimum' to be merely a local minimizer.
Note that, when function $Q$ is strictly convex, the unique minimizer is the global minimizer,
therefore the relaxation of optimal does not affect convex problems, where $x \opt$  is still the true global minimizer.

We assume the unknown $x \in \real^{N}$ can be divided into three blocks: $x_0 \in \real^{N_0}$, $x_1 \in \real^{N_1}$ and $x_2 \in \real^{ N_2}$ with $N = N_0 + N_1 + N_2$.
Then the objective function $Q$ in \eqnref{eqn:bsum_bsum_cost} can be rewritten using the expression of the three blocks, where the problem is
\begin{align}\label{eqn:bsum_bsum_cost2}
	\{ x_0\opt, x_1\opt , x_2\opt \} = 	& \arg \min_{x\in \real^{N}}  Q(x_0, x_1, x_2),  \text{ where } x = \begin{bmatrix} x_0 \\ x_1 \\ x_2 \end{bmatrix},
\end{align}
and if we can find a continuous function $G: \real^{N} \times \real^{N} \to \real$, which is the block-wise upper-bound of $Q$,
defined as
\begin{subequations}\label{eqn:bsum_block_condition}
\begin{align}
	&	G( ( x_0, x_1, x_2 ), ( v_0, x_1, x_2) ) \ge Q(x_0, x_1, x_2), \quad \text{ for all } x \text{ and } v_0 \in \real^{N_0}, \\
	&	G( ( x_0, x_1, x_2 ), ( x_0, v_1, x_2) ) \ge Q(x_0, x_1, x_2), \quad \text{ for all } x \text{ and } v_1 \in \real^{N_1}, \\
	&	G( ( x_0, x_1, x_2 ), ( x_0, x_1, v_2) ) \ge Q(x_0, x_1, x_2), \quad \text{ for all } x \text{ and } v_2 \in \real^{N_2}, \\
	&	G( ( x_0, x_1, x_2 ), ( x_0, x_1, x_2) ) =   Q(x_0, x_1, x_2), \quad \text{ for all } x,
\end{align}
\end{subequations}
then problem \eqnref{eqn:bsum_bsum_cost2} can be solved by BSUM by the algorithm illustrated in \cite[Figure~2]{opt_Razaviyayn_2013}.
For the special case of three blocks, we can write the algorithm explicitly as the steps in \tabref{alg:bsum}.

The condition \eqnref{eqn:bsum_block_condition} implies all the assumptions listed in \cite[Assumption 2]{opt_Razaviyayn_2013},
which ensures that the algorithm in \tabref{alg:bsum} converges,
and the mathematical proof of convergence can be found in Section~IV of \cite{opt_Razaviyayn_2013}.
It is worth noting that the convergence behavior of BSUM does not rely on the convexity of the problem.

\begin{table}[htbp]
\caption{Block successive upper-bound minimization (BSUM) algorithm with three blocks.}
\label{alg:bsum}
\begin{subequations}
	\begin{empheq}[box=\fbox]{align*}
		& \text{initial:} ~x \in \real^{N}	\\
		& \text{repeat:} \\
		&  	\quad x_0 = \arg \min_{u_0} G( ( u_0, x_1, x_2 ), ( x_0, x_1, x_2) ) \\
		&  	\quad x_1 = \arg \min_{u_1} G( ( x_0, u_1, x_2 ), ( x_0, x_1, x_2) ) \\
		&  	\quad x_2 = \arg \min_{u_2} G( ( x_0, x_1, u_2 ), ( x_0, x_1, x_2) ) \\
		& \text{end}\\
		& \text{return: } x = \{ x_0, x_1, x_2\}
	\end{empheq}
\end{subequations}
\end{table}

\section{SALMA}

Split augmented Lagrangian shrinkage algorithm (SALSA) \cite{Afonso_2010_TIP_SALSA} is a recently developed method for solving convex optimization problems.
As an efficient $\ell_1$-norm regularized convex problem solver, it has been widely used to solve various signal processing problems.
SALSA adopts variable splitting technique and the concept of alternating direction method of multipliers (ADMM)
\cite{Boyd_2011_admm, opt_Eckstein_1992}, achieving a very simply structured and efficient iterative algorithm.
In this section, we propose an algorithm formulated very similar to SALSA.
However, in contrast to SALSA, the proposed method converges when used to solve non-convex problems.
Because we use the concept of majorization-minimization (MM) in the proposed algorithm,
it is termed split augmented Lagrangian majorization-minimization algorithm (SALMA),
and can be stated as the following theorem.

\begin{theorem}\label{the:salma}
For optimization problem
\begin{align}\label{eqn:bsum_problem}
	 x\opt  = 	& \arg \min_{x \in \real^{N}}F_0(x) + F_1(x),
\end{align}
where function $F_0 : \real^{N} \to \real$ and $ F_1 : \real^{N} \to \real$ are both finite and continuously differentiable,
if a function $G_1: \real^{N} \times \real^{N} \to \real$ majorizes $F_1$, i.e., satisfying
\begin{align}\label{eqn:fs_the_condition}
	& G_1(u,v) 	\ge F_1(u),\quad \text{ for all } u, v \in \real^{N},\\
	& G_1(u,u) 	= 	F_1(u),
\end{align}
can be found, and it enables the following iterative algorithm
\begin{subequations}\label{eqn:bsum_salma_0}
\begin{align}
	u^{(i+1)}	& = \arg \min_{u} 	G_1(u, u^{(i)}) +  \frac{\mu}{2} \norm{u-x-d}_2^2, 	\label{eqn:bsum_salma_0_a}	 \\
	x^{(i+1)}	& = \arg \min_{x}	F_0(x) + \frac{\mu}{2} \norm{u-x-d}_2^2,  	\\
	d^{(i+1)}	& = d^{(i)} - (u-x)	\label{eqn:bsum_salma_0_c}											
\end{align}
\end{subequations}
having unique solution at each step,
then the algorithm \eqnref{eqn:bsum_salma_0} will converge, and converges to a local minimizer of problem \eqnref{eqn:bsum_problem}.
\end{theorem}

The proof of the above theorem in detail is given in \appref{app:proof}.
Note that the algorithm \eqnref{eqn:bsum_salma_0} has two steps identical to SALSA,
and the difference is in the sub-problem \eqnref{eqn:bsum_salma_0_a}, where the cost function is based on a continuously differentiable majorizer of $F_1$.
In contrast to SALSA, the proposed method will converge when the objective function is not convex, because it is actually a special case of BSUM.

In other words, the proposed algorithm utilizes a Gauss-Seidel procedure to solve a majorization-minimization (MM) problem.
This leads us to term the algorithm SALMA (split augmented Lagrangian majorization-minimization algorithm),
which combines SALSA and majorization-minimization methods.

\section{Proposed method}

\subsection{Smoothed penalty function}

\begin{figure}[t]
\centering
	\includegraphics[scale = \figurescalesmall] {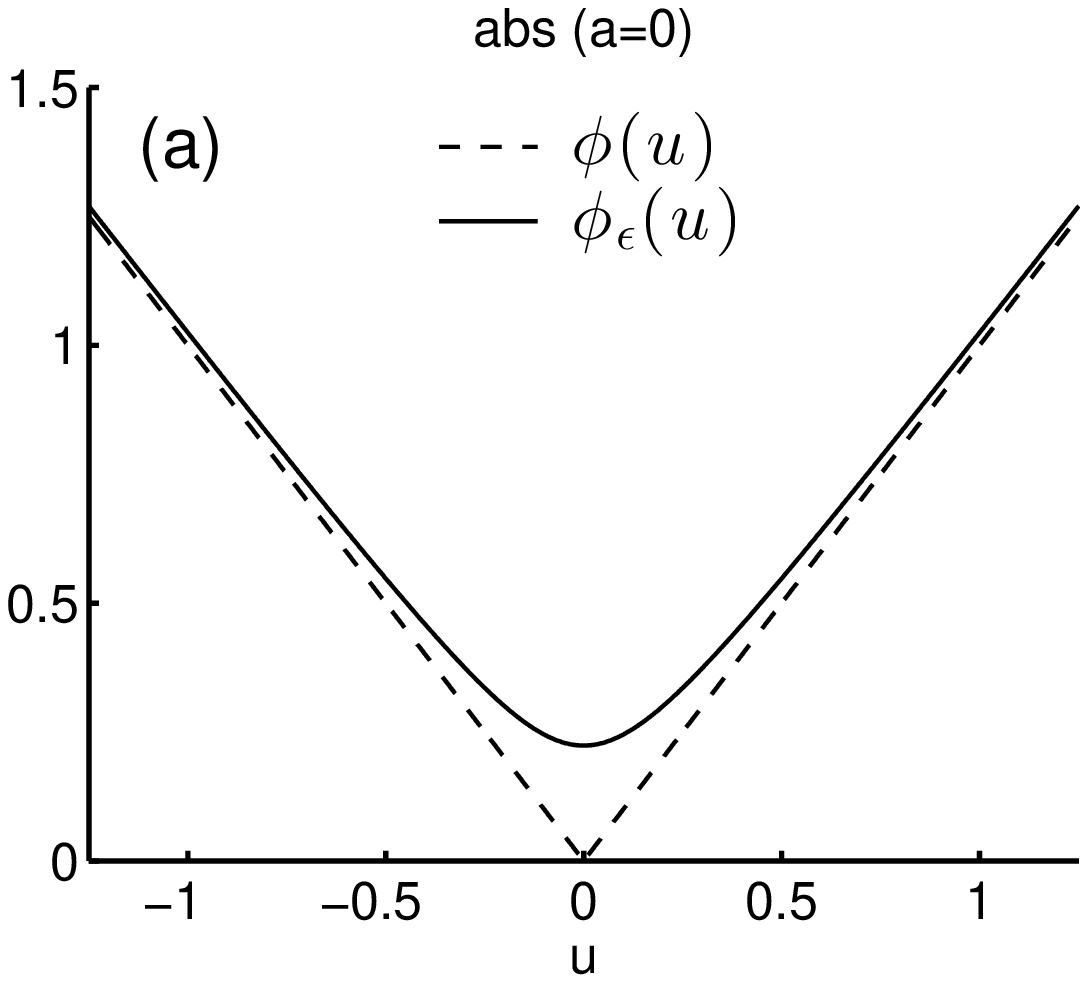}	
	\includegraphics[scale = \figurescalesmall] {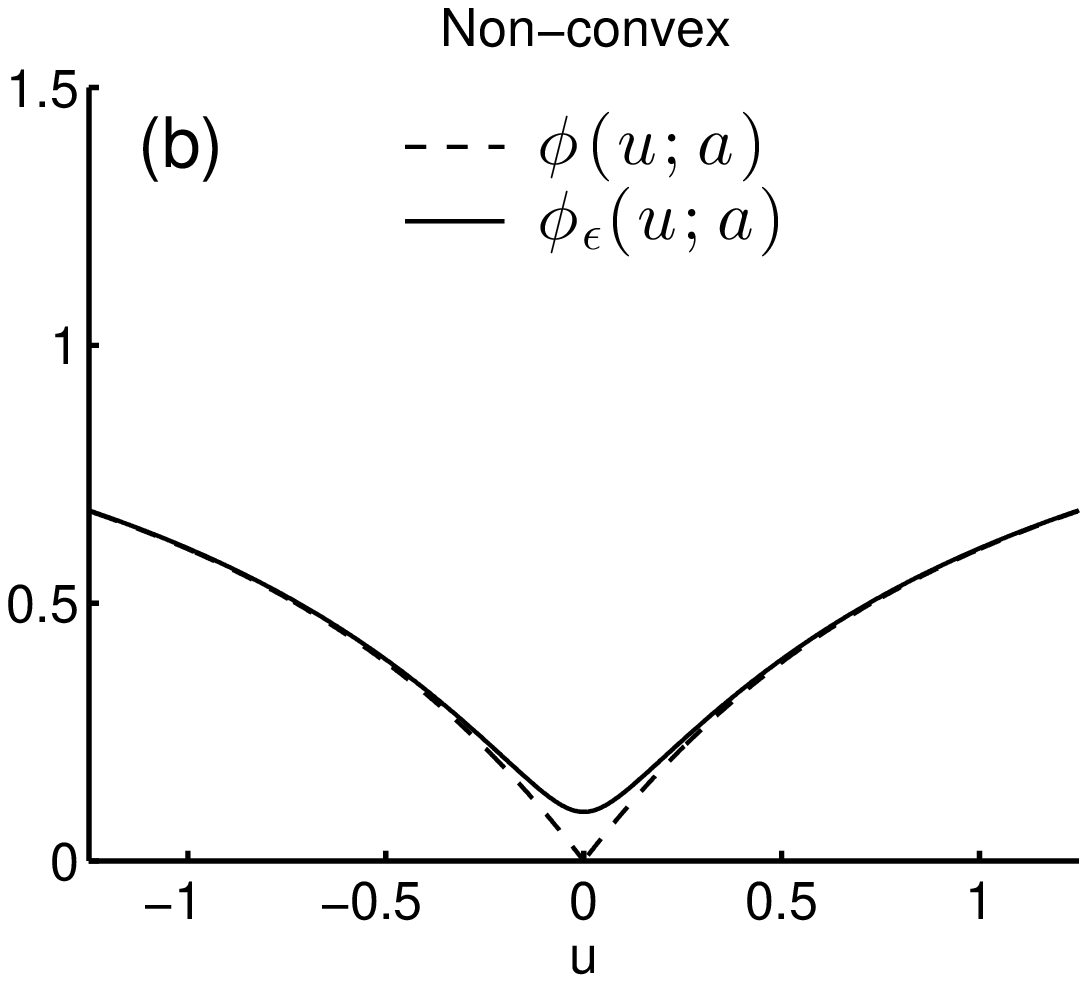}	
	\\	
	\includegraphics[scale = \figurescalesmall] {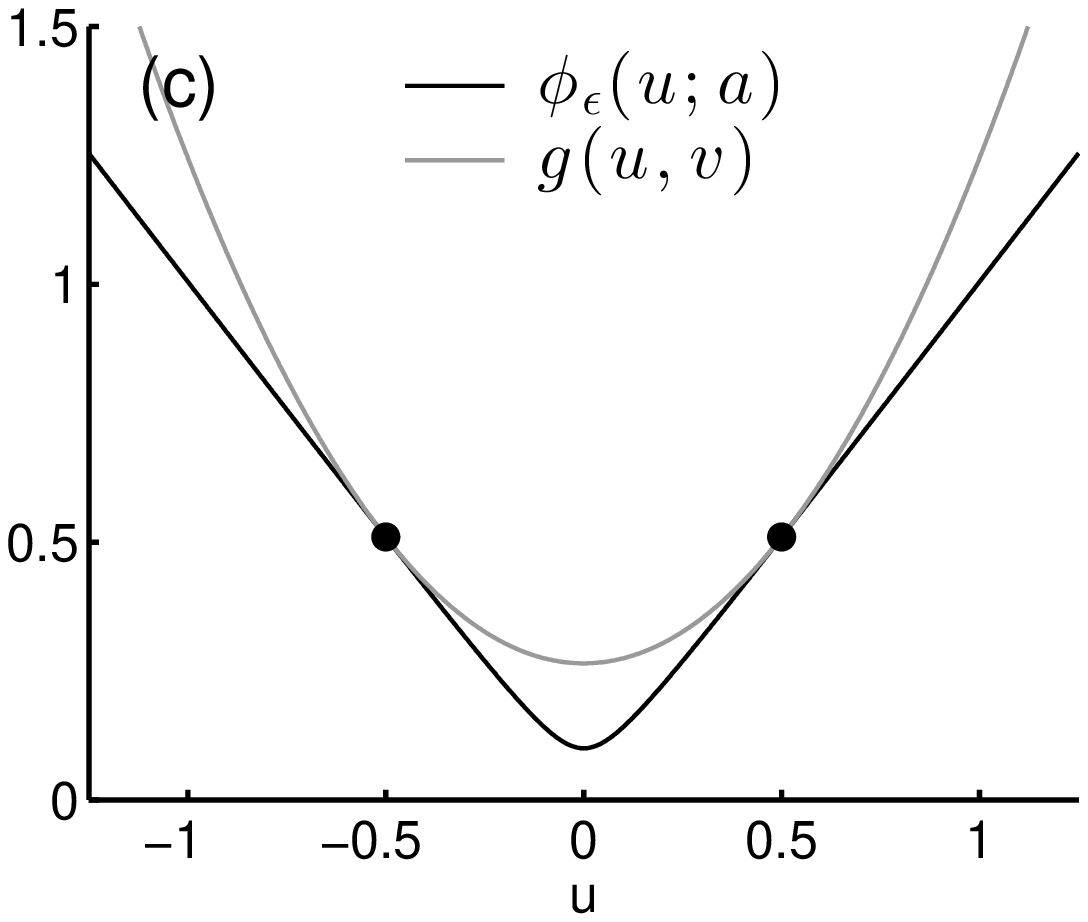}	
	\includegraphics[scale = \figurescalesmall] {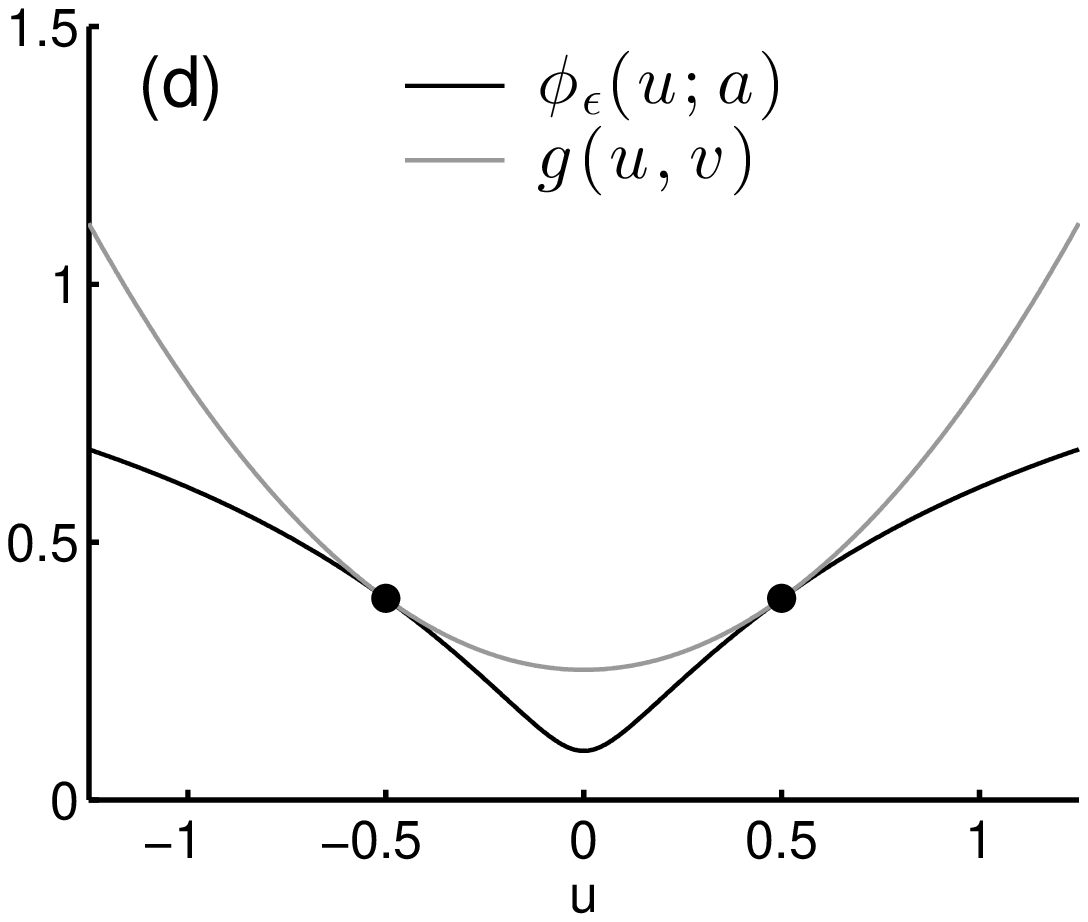}			
	\caption{%
		(a) $\ell_1$-norm penalty function and its smoothed version, and
		(b) arctangent penalty function and its smoothed version, and
		(c) Majorizer (gray) of the smoothed penalty function in (a).
		(d) Majorizer (gray) of the smoothed penalty function in (b).
		For all the figures we set $\eps = 0.01, v = 0.5, a = 1$.	}
	\label{fig:fs_Example_0_penalty}
\end{figure}

In this work, we use smoothed penalty function $\phie : \real \to \real_{+}$ to induce sparsity,
which is a smoothed version of $\phi : \real \to \real_{+}$ in \tabref{tab:etea_penalty}.
It is defined as:
\begin{align}\label{eqn:fs_phie}
	\phie(u ; a) : = \phi( \sqrt{ u^2 +\eps } ; a ), \quad \eps > 0,
\end{align}
where $\phi$ is the non-smooth penalty function satisfies the following properties:
\begin{enumerate}
\item
	$\phi(u;a)$ is continuous on $\real$.
\item
	$\phi(u;a)$ is twice continuously differentiable on $\mathbb{R} \backslash \{0\}$.
\item
	$\phi(u;a)$ is even symmetric:  $\phi(u) = \phi(-u)$.
\item
	$\phi(u;a)$ is increasing and concave on $\real_{+}$.
\item
	$\phi(u;a) = \abs{x}$, when $a = 0$.
\end{enumerate}
Several typical examples of $\phi$ are given in \tabref{tab:etea_penalty}.
\figref{fig:fs_Example_0_penalty}(a-b) gives two specific examples of $\phi$ in dashed line,
and corresponding smoothed versions with $\eps = 0.01$ in solid line.

In this work, we use the smoothed penalty function $\phie$ instead of $\phi$, because after the smoothing, the function is continuously differentiable on $u \in \real$,
and we can easily find the corresponding majorizer,
\begin{align}\label{eqn:fs_g}
	g( u , v ) := \frac{u^2}{ 2 \psi(v) }  - \underbrace{ \bigg( \frac{v^2}{2 \psi(v)} - \phie(v;a) \bigg) }_{\text{only depends on $v$}}.
\end{align}
where $\psi(v) := v / \phie'(v ; a)$, and $g(u , v )$ is quadratic in $u$.
In Ref.~\cite{Ding_2015_SP}, the same smoothed penalty functions are used
and a detailed proof is given to show that when $\phi$ satisfies the above 5 conditions,
the majorizer \eqnref{eqn:fs_g} satisfies
\begin{subequations}\label{eqn:fs_g_mm}
\begin{align}
	&g(u,v) 	\ge \phie(u;a),\quad \text{ for all } u, v,\\
	&g(u,u) 	= 	\phie(u;a).
\end{align}
\end{subequations}

\figref{fig:fs_Example_0_penalty}(c-d) illustrates the majorizer $g$ of two penalty functions when $ v = 0.5 $.
Note that $g$ never goes to extrema (i.e. $ \pm \infty $) if $u$ and $v$ are both finite.
\tabref{tab:etea_penalty} provides four examples of smoothed penalty functions.
Among all these functions, only the first one (`abs') is convex, and the rest are all non-convex.
In \eqnref{eqn:fs_phie}, $\phi$ is not differentiable at $0$, but $\phie$ is differentiable,
and when $\eps \to 0$, function $\phie \approx \phi$, but preserving the differentiability at the origin.
In practice, we recommend to set $\eps$ very small, e.g. $10^{-8}$.

\begin{table*}[t]
\caption{Sparsity-promoting penalty functions.}
\begin{center}
	\scalebox{0.80}{
		\begin{tabular}{@{} c l l l l@{}}
		\toprule
		\textbf{Penalty}	 &$\phi(u;a)$ 	&$\phie(u;a)$  &$\psi(u) = u / \phie'(u;a) \quad $ \\[0.4em]
		\midrule
			abs ($a= 0$)	 	& $\abs{u}$						
	 				& $\sqrt{ u^2 + \eps }$		
	 				& $\sqrt{ u^2 + \eps }$ \\[0.8em]	
			log	 	&$\ds\frac{1}{a} \log ( 1 + a |u| )$
	 				&$\ds\frac{1}{a} \log ( 1 + a \sqrt{u^2+\eps} )$
	 				&$\ds\sqrt{ u^2 + \eps } \left( 1 + a \sqrt{ u^2 + \eps } \right)$	\\[0.8em]
			rat	 	&$\ds\frac{\abs{u}}{ 1 + a\abs{u} /2 }$
	 				&$\ds\frac{ \sqrt{ u^2 + \eps} }{ 1 + a \sqrt{ u^2 + \eps } /2 }$
	 				&$\ds\sqrt{ u^2 + \eps } \left( 1 + a \sqrt{ u^2 + \eps } /2\right)^2$	\\[0.8em]
			atan 	&$\ds\frac{2}{a \sqrt{3}} \left( \tan^{-1} \left( \frac{1+2 a |u|} {\sqrt{3}} \right) - \frac{\pi}{6}\right)$
	 				&$\ds\frac{2}{a \sqrt{3}} \left( \tan^{-1} \left( \frac{1+2 a \sqrt{u^2+\eps} } {\sqrt{3}} \right) - \frac{\pi}{6} \right)$
	 				&$\ds\sqrt{ u^2 + \eps } \left( 1 + a \sqrt{ u^2 + \eps } + a^2 (u^2 + \eps) \right)$	\\[0.8em]
		\bottomrule
		\end{tabular}
	}
	\end{center}
	\label{tab:etea_penalty}	
\end{table*}

\subsection{Problem definition}

To estimate a sequence of oscillatory transients modeled by \eqnref{eqn:fs_model},
through an optimization problem prototyped by \eqnref{eq:optimization},
we further formulate the optimization problem explicitly as
\begin{align}\label{eqn:fs_cost}
	c \opt = \arg \min_{c}
	\Bigg\{ P(c) =
	\half \norm{y - A c}_2^2 + \lam \sum_{m_1,m_2}  \phi_{\eps} \big( \bnorm(c, B, m_1,m_2) ; a \big)
	\Bigg\}
\end{align}
where
$ y \in \real^{N}$,
$ c \in \mathbb{C}^{M_1 \times M_2} $ are STFT coefficients,
and the matrix
$A : \mathbb{C}^{M_1 \times M_2} \to \mathbb{C}^{N}$ here denotes the inverse short-time Fourier transform operator.
When a local optimal of problem \eqnref{eqn:fs_cost} is obtained,
component $x$ (periodic sequence of oscillatory transients) in signal model \eqnref{eqn:fs_model} is given by
\begin{align}
	\hat x = A c\opt.
\end{align}
In problem \eqnref{eqn:fs_cost}, we defined the regularizer by function
$\bnorm : \complex^{M_1 \times M_2} \times \real^{K_1 \times K_2} \times \mathbb{Z} \times \mathbb{Z} \to \real $,
\begin{align}
	\bnorm(c,B, m_1,m_2)
		& = \bigg[  \sum_{k_1=0}^{K_1 - 1} \sum_{k_2=0}^{K_2 - 1}  [B]_{k_1,k_2} [c]_{m_1 + k_1, m_2 + k_2}^2  \bigg]^{1/2},	
\end{align}
which is similar to a 2-D convolution.
Note that $B \in \real^{K_1 \times K_2}$ is a two-dimensional binary-valued mask describing the group structure,
which is to be determined beforehand by the time-frequency information of $x$ and the STFT.

We use $\odot$ to denote element-wise multiplication, so the function $\bnorm$ can be written as a weighted norm of a segment of STFT coefficient $c$,
\begin{align}\label{eqn:fs_bnorm}
	\bnorm(c,B, m_1,m_2) : = \norm{ B \odot S(c, m_1, m_2) }_2.
\end{align}
Here $S(c,m_1, m_2)$ denotes a $K_1 \times K_2$ segment of matrix $c \in \complex^{M_1 \times M_2}$,
where the first element (up and left) is $[c]_{m_1,m_2}$.
Moreover, since $c$ is a complex-valued array, where $ c = c_r + \ii c_i $, it follows that,
\begin{align}
	\norm{ B \odot S(c, m_1, m_2) }_2^2 = \norm{ B \odot S(c_r, m_1, m_2) }_2^2 + \norm{ B \odot S(c_i, m_1, m_2) }_2^2
\end{align}
where $ S(c_r, m_1, m_2), S(c_i, m_1, m_2) \in \real^{K_1 \times K_2} $ are the corresponding arrays of the real and imaginary parts of $S(c, m_1, m_2)$.
Then, in implementation, the real and imaginary parts of the problem can be calculated separately as real values.

\begin{figure}[h]
\centering
	\includegraphics[scale = 0.15] {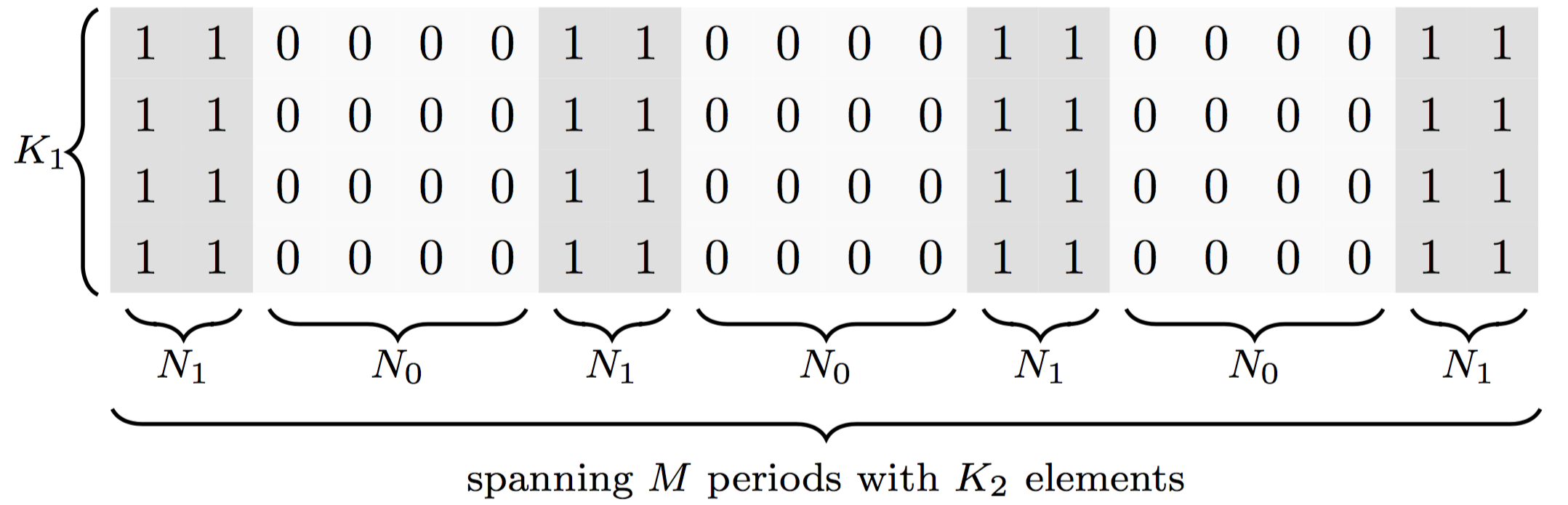}
\caption{Example of binary block $B$ when $M = 4$, $K_1 = 4$, $N_0 = 4$ and $N_1 = 2$.}
\label{fig:fs_B}
\end{figure}

For problem \eqnref{eqn:fs_cost}, in order to capture the known periodicity of $x$ as modeled in \eqnref{eqn:fs_model},
we further define binary weighting vectors $b_1 \in \{ 0,1 \}^{K_1}$ and $b_2 \in \{ 0,1 \}^{K_2}$ in the following structure,
\begin{subequations}\label{eqn:fd_preoid_b}
\begin{align}
		b_1 & = \mathbf{1}{(1, K_1)}  = [\underbrace{ 1, 1, \ldots , 1 }_{K_{1}}] \\
		b_2 & = [ \mathbf{1}(1, N_1), \ \mathbf{0}(1, N_0), \ \mathbf{1}(1, N_1), \ \mathbf{0}(1, N_0), \	\ldots, \ \mathbf{1}(1, N_1) ] \nonumber \\
		& = [
			\ 	\underbrace{ 1, 1, \ldots , 1 }_{N_{1}},
			\	\underbrace{ 0, 0, \ldots , 0 }_{N_{0}},
			\	\underbrace{ 1, 1, \ldots , 1 }_{N_{1}},
			\ 	\underbrace{ 0, 0, \ldots , 0 }_{N_{0}} ,
			\	\ldots, \	
			\ 	\underbrace{ 1, 1, \ldots , 1 }_{N_{1}} \
			],
\end{align}
\end{subequations}
where  $b_2$ captures the feature periodicity with a periodic binary structure,
so that the binary block $B$ spanning $M$ periods can be written explicitly as
\begin{align}\label{eqn:fs_block}
	B & = b_1^{\tp} \times b_2
	 = [ \ \underbrace{  \mathbf{1}(K_1, N_1), \ \mathbf{0}(K_1, N_0), \ \mathbf{1}(K_1, N_1), \ \mathbf{0}(K_1, N_0), \	
	 	\ldots, \ \mathbf{1}(K_1, N_1) }_{\text{spanning } M \text{ periods}} \ ],
\end{align}
which is concatenating of $M$ all-$1$'s blocks and $M-1$ all-$0$'s blocks intervally.
For instance, when $M = 4$, $K_1 = 4$, $N_0 = 4$ and $N_1 = 2$, the block $B$ is a binary grid illustrated in \figref{fig:fs_B}.

From \eqnref{eqn:fs_block} we can derive that the two-dimensional binary-weight block $B \in \real^{K_1 \times K_2}$ has a size determined by $K_1$ and
\begin{align}\label{eqn:fs_block_K}
	K_2 & = (N_{0} + N_{1}) \times (M-1) + N_{1}.
\end{align}
Using $B$ in the objective function of \eqnref{eqn:fs_cost},
the regularization term groups a 2-D structure of STFT coefficients with the respect to the periodicity information of $x$ in \eqnref{eqn:fs_model}.

\subsection{Algorithm derivation}

To solve the optimization problem \eqnref{eqn:fs_cost}, we split the variable $c$ by introducing an extra equality constraint,
and solve the equivalent problem
\begin{subequations}\label{eqn:fs_cost_2}
\begin{align}
	\{ c\opt,  u\opt \} =
	& \arg \min_{c, u} P_0(c) + P_1(u)  \\
	& \text{ subject to } c - u = 0
\end{align}
\end{subequations}
where $P_0 : \complex^{M_1 \times M_2} \to \real $ and $P_1 : \complex^{M_1 \times M_2} \to \real $ are defined as,
\begin{subequations}\label{eqn:fs_vs}
\begin{align}
	P_0(c) & : = \half \norm{y - A c}_2^2 \\
	P_1(u) & : = \lam \sum_{m_1}\sum_{m_2}  \phi_{\eps} \big( \bnorm(u,B, m_1,m_2) ; a \big),
\end{align}
\end{subequations}
which leads to a same solution by $ c = u $ to problem \eqnref{eqn:fs_cost}.

The majorizer of $P_1$ in \eqnref{eqn:fs_cost_2} can be found by $g(u,v)$ in \eqnref{eqn:fs_g} as,
\begin{subequations} \label{eqn:fs_Guv}
\begin{align}
	G_1(u,v)
		& = \lam \sum_{m_1}\sum_{m_2}  g\big( \bnorm(u,B, m_1,m_2) , \bnorm(v,B, m_1,m_2)  \big) \\
		& = \lam \sum_{m_1}\sum_{m_2}  g\big( \norm{ B \odot S(u,m_1,m_2)}_2 , \norm{ B \odot S(v,m_1,m_2)}_2  \big) \\
		& = \frac{\lam}{2} \sum_{m_1}\sum_{m_2}  [r(v)]_{m_1,m_2}  [u]_{m_1,m_2}^2 + C
\end{align}
\end{subequations}
where $C$ is a constant, and $ r(v) \in \real^{ M_1 \times M_2 } $ is a matrix having an explicit form
\begin{align}\label{eqn:fs_rmn}
	[r(v)]_{m_1,m_2}  = \sum_{k_1 = 0} ^{K_1-1} \sum_{k_2 = 0} ^{K_2-1} \frac{ [B]_{k_1,k_2}^2 }{ \psi( \norm{B \odot S( v, m_1-k_1, m_2-k_2 )}_2  ) }.
\end{align}
Then using SALMA (\theref{the:salma}), problem \eqnref{eqn:fs_cost} can be solved iteratively by
\begin{subequations}\label{eqn:fs_salma_1}
\begin{align}
	u^{(i+1)}	& = \arg \min_{u} 	G_1(u, u^{(i)}) +  \frac{\mu}{2} \norm{u-c-d}_2^2 	\label{eqn:fs_salma_1_a}  \\
	c^{(i+1)}	& = \arg \min_{c}	P_0(c) + \frac{\mu}{2} \norm{u-c-d}_2^2 \label{eqn:fs_salma_1_b} 	\\
	d^{(i+1)}	& = d^{(i)} - (u-c).										
\end{align}
\end{subequations}

\subsubsection{Sub-problem \eqnref{eqn:fs_salma_1_a} }
The step \eqnref{eqn:fs_salma_1_a} can be written as
\begin{align}
	u	& = \arg \min_{u} 	G_1(u, v) +  \frac{\mu}{2} \norm{u-c-d}_2^2
\end{align}
where $G_1(u, v)$ is given by \eqnref{eqn:fs_Guv}, therefore the problem has a more explicit formulation
\begin{align}
	u	& = \arg \min_{u} \frac{\mu}{2} \norm{u-(c+d)}_2^2 + \frac{\lam}{2} \sum_{m_1}\sum_{m_2} [r(v)]_{m_1,m_2} [u]_{m_1,m_2}^2,
\end{align}
which is a least square problem with explicit solution
\begin{align}\label{eqn:fs_salma_1_a_solution_2}
	u	& = ( c + d ) ./ \Big[ 1 + \frac{\lam [r(v)] }{\mu}  \Big],
\end{align}
where $ ./  $ denotes the point-wise division.

\subsubsection{Sub-problem \eqnref{eqn:fs_salma_1_b} }
The sub-problem \eqnref{eqn:fs_salma_1_b} can be written as
\begin{align}
	c	& = \arg \min_{c} 	\half \norm{y - A c}_2^2+  \frac{\mu}{2} \norm{u-c-d}_2^2
\end{align}
which is also a least-square problem, and it has an explicit solution
\begin{align}\label{eqn:fs_salma_1_b_solution_1}
	c = \Big[ A^{\ct} A + \mu I \Big]^{-1} \Big[ A^{\ct} y + \mu( u-d ) \Big].
\end{align}
The operator $A^{\ct}$ is the complex conjugate transpose (Hermitian) of $A$.
Note that to compute $c$ by \eqnref{eqn:fs_salma_1_b_solution_1} we need to solve an inverse system $\Big[ A^{\ct} A + \mu I \Big]^{-1}$.
To overcome the computational cost,
we use the matrix inverse lemma (see the formulation in \appref{app:A}),
\begin{subequations}
\begin{align}\label{eqn:fs_inverse_1}
	\Big[ A^{\ct} A + \mu I \Big]^{-1}
		& = \frac{1}{\mu} \bigg\{ I  -  A^{\ct} \Big[ \mu I + A A^{\ct} \Big]^{-1} A \bigg\} \\
		& = \frac{1}{\mu} \Big[ I  -  \frac{1}{\mu+1}A^{\ct} A \Big],
\end{align}
\end{subequations}
where since the operators $A^{\ct}$ and $A$ here are the forward and inverse STFT, where $x = A A^{\ct} x$,
then the property
\begin{align}\label{eqn:fs_property}
	A A^{\ct} = I
\end{align}
holds.
As a consequence, equation~\eqnref{eqn:fs_salma_1_b_solution_1} can be written as,
\begin{subequations}\label{eqn:fs_salma_1_b_solution_2}
\begin{align}
	c	& = \frac{1}{\mu} \Big[ I  -  \frac{1}{\mu+1}A^{\ct} A \Big] \Big[ A^{\ct} y + \mu( u-d ) \Big] \\
		& = \frac{1}{\mu}A^{\ct} y + ( u-d )  -  \frac{1}{\mu(\mu+1)} A^{\ct} y - \frac{1}{1+\mu} A^{\ct} A( u-d ) \\
		& = ( u-d ) + \frac{1}{\mu+1} A^{\ct} \Big[y - A( u-d ) \Big],
\end{align}
\end{subequations}
which is equivalent to \eqnref{eqn:fs_salma_1_b_solution_1},
but does not require solving a linear system in \eqnref{eqn:fs_salma_1_b_solution_1}.

Using the results derived in \eqnref{eqn:fs_salma_1_a_solution_2} and \eqnref{eqn:fs_salma_1_b_solution_2},
which solve the problems \eqnref{eqn:fs_salma_1_a} and \eqnref{eqn:fs_salma_1_b} respectively,
the algorithm \eqnref{eqn:fs_salma_1} has explicit solutions of each step,
\begin{subequations}\label{eqn:fs_salma_2}
\begin{align}
	u^{(i+1)}	& = ( c + d ) ./ \Big[ 1 + \frac{\lam r( u^{(i)} ) }{\mu}   \Big] 	 \\
	c^{(i+1)}	& = ( u-d ) + \frac{1}{\mu+1} A^{\ct} \Big[y - A( u-d ) \Big] 	\\
	d^{(i+1)}	& = d^{(i)} - (u-c)	.								
\end{align}
\end{subequations}
The entire algorithm to solve problem \eqnref{eqn:fs_cost} is summarized in \tabref{alg:main}.
Note that although the variables $u, c, d \in \complex^{M_1 \times M_2 }$ in the algorithm are all 2-D complex-valued matrices,
in fact this does not affect the feasibility of SALMA.
In each step of the algorithm (\tabref{alg:main}), the real and imaginary parts can be computed parallelly and separately, using real-valued expressions.
More specifically, it is equivalent to implement the algorithm using 3-D matrices on $\real^{M_1 \times M_2 \times 2}$,
and each step leads to a same result.
Hence, SALMA (\theref{the:salma}) defined in real domain,
is still valid to solve the problem \eqnref{eqn:fs_cost} (see \appref{app:B} for more details).

\begin{table}[htbp]
\caption{Solving problem \eqnref{eqn:fs_cost} with SALMA.}
\label{alg:main}
\begin{subequations}
	\begin{empheq}[box=\fbox]{align*}
		& \text{initial:} ~c, u, d \in \complex^{ M_1 \times M_2}, \mu > 0, \\
		& \text{repeat:} \\
		& 	\quad [r]_{m_1,m_2} = \sum_{k_1} \sum_{k_2} \frac{ [B]_{k_1,k_2}^2 }{ \psi( \norm{B \odot S( u, m_1-k_1, m_2-k_2 )}_2  ) } \\[0.4em]
		&  	\quad u = ( c + d ) ./ \Big[ 1 + \frac{\lam }{\mu} r   \Big]\\
		&  	\quad c = ( u-d ) + \frac{1}{\mu+1} A^{\ct} \Big[y - A( u-d ) \Big]  \\
		&  	\quad d = d - (u-c) \\
		& \text{end}\\
		& \text{return: } c, \ x = Ac.
	\end{empheq}
\end{subequations}
\end{table}

\section{Simulated data example}

\subsection{Example 1}

\begin{figure}[t]
\centering
	\includegraphics[scale = \figurescale] {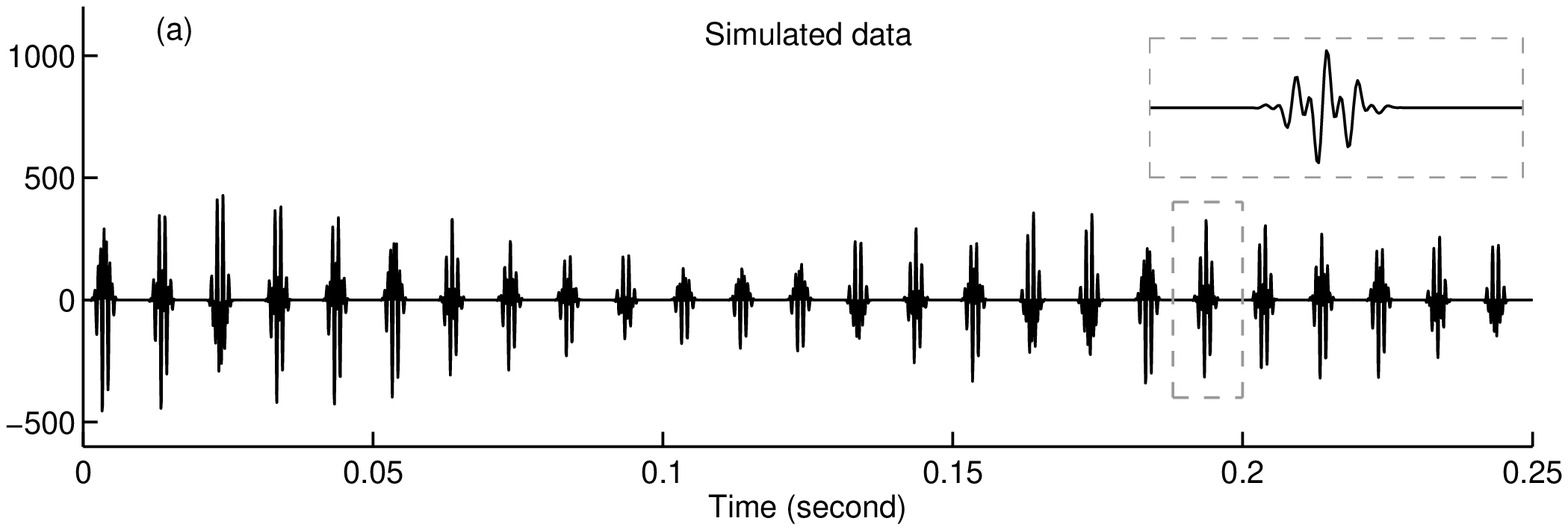}		
	\\
	\includegraphics[scale = \figurescale] {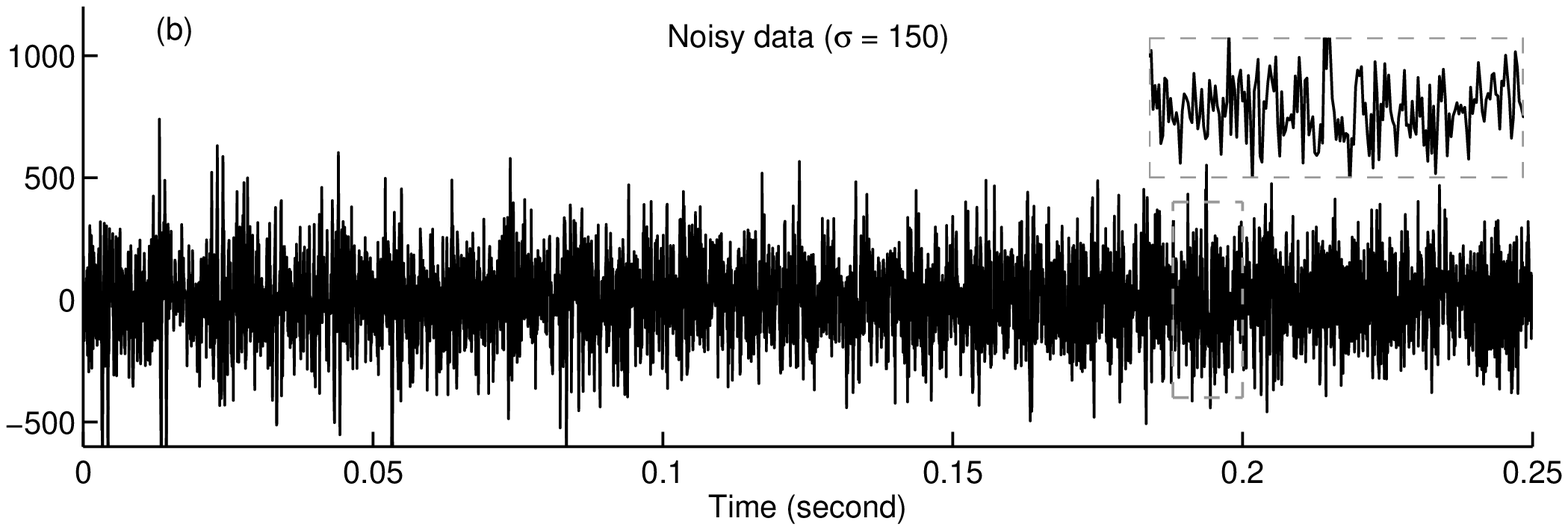}		
	\caption{(a) Simulated signal, and (b) noisy observation.}
	\label{fig:fs_Example_1_y}
\end{figure}
\begin{figure}[t]
\centering
	\includegraphics[scale = \figurescale] {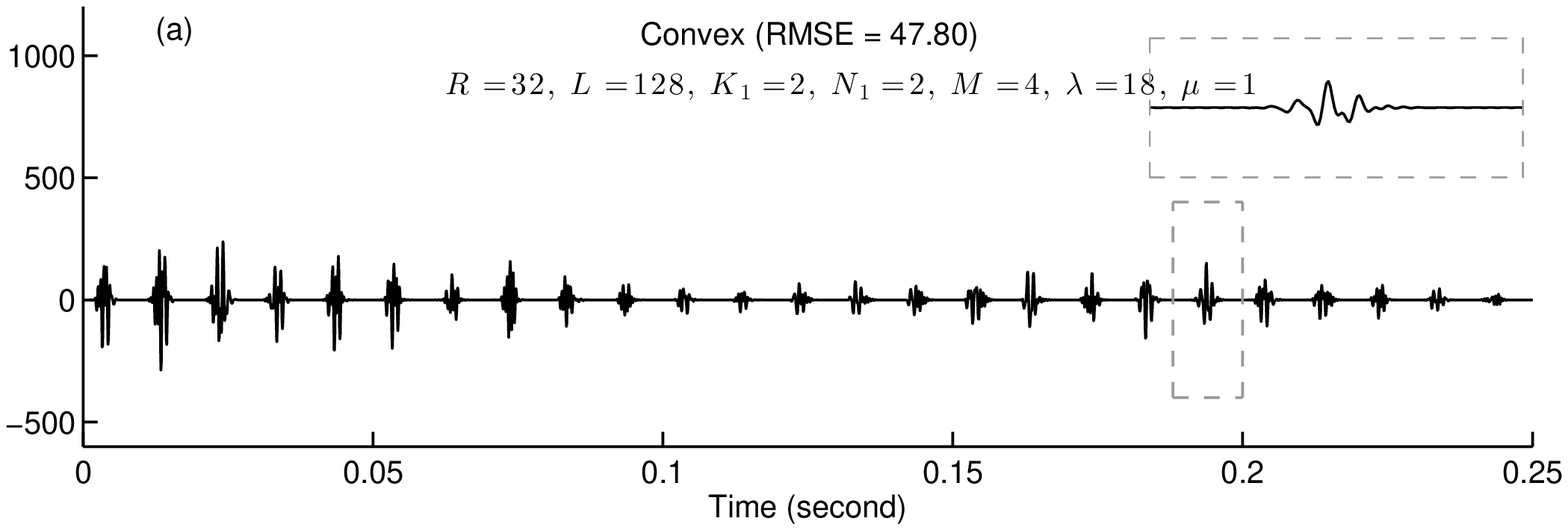}	
	\\
	\includegraphics[scale = \figurescale] {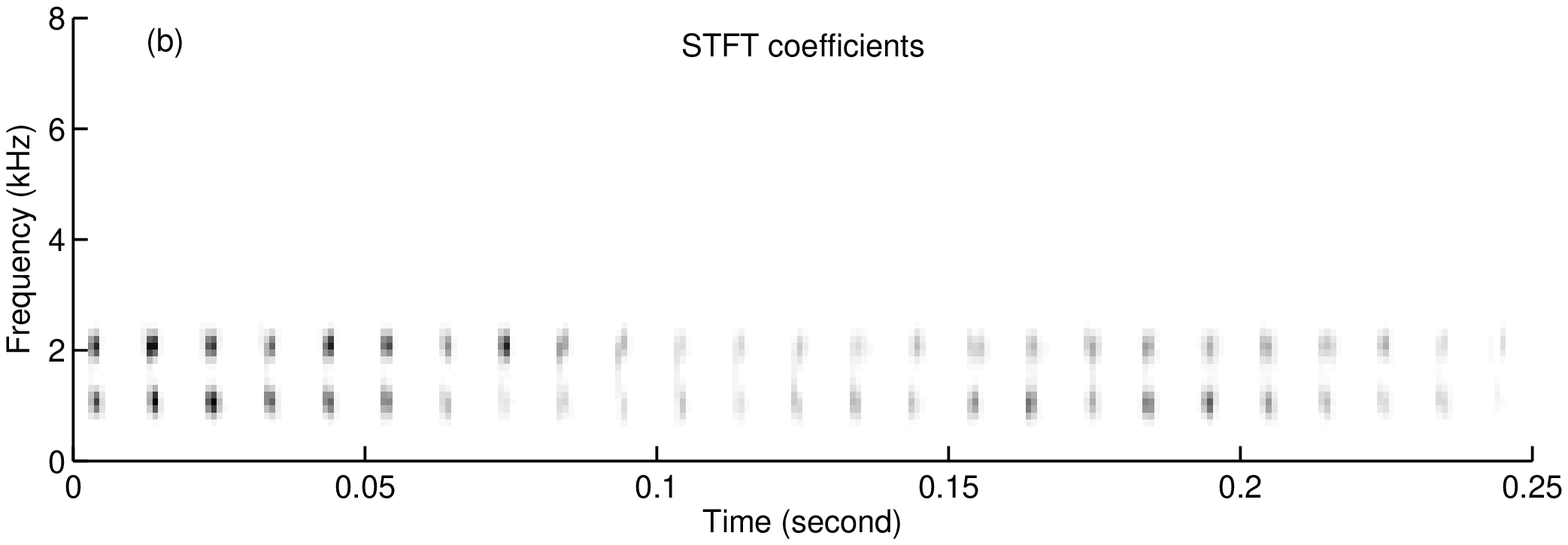}
	\\
	\includegraphics[scale = \figurescale] {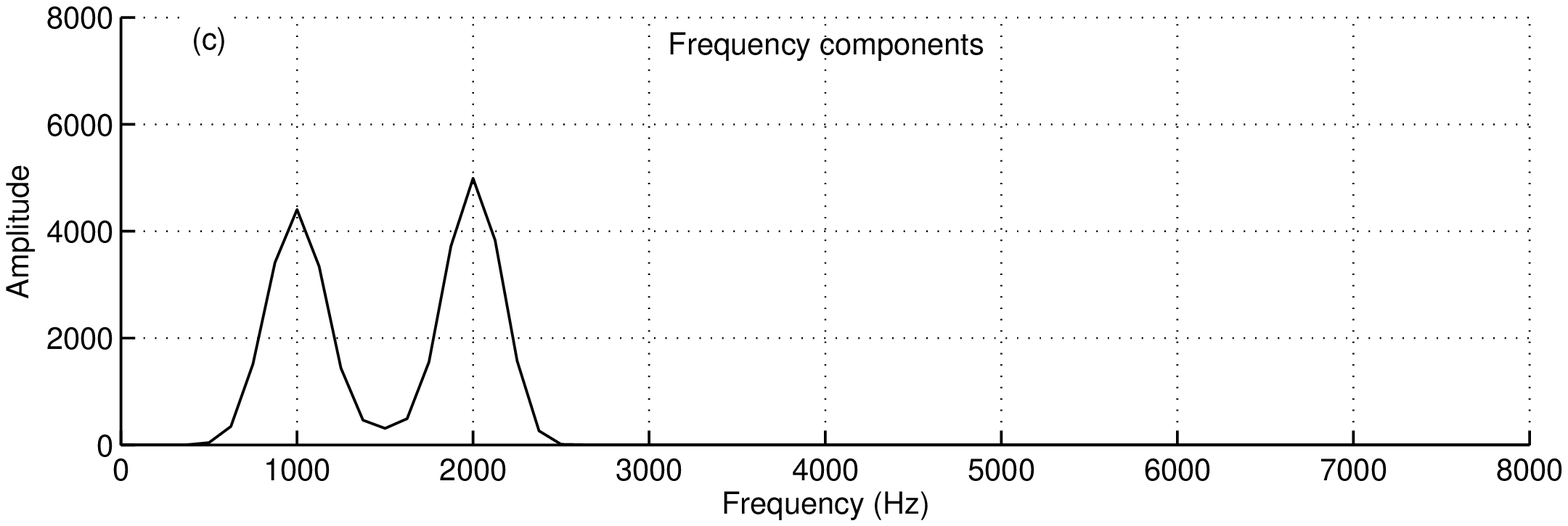}
	\\
	\includegraphics[scale = \figurescale] {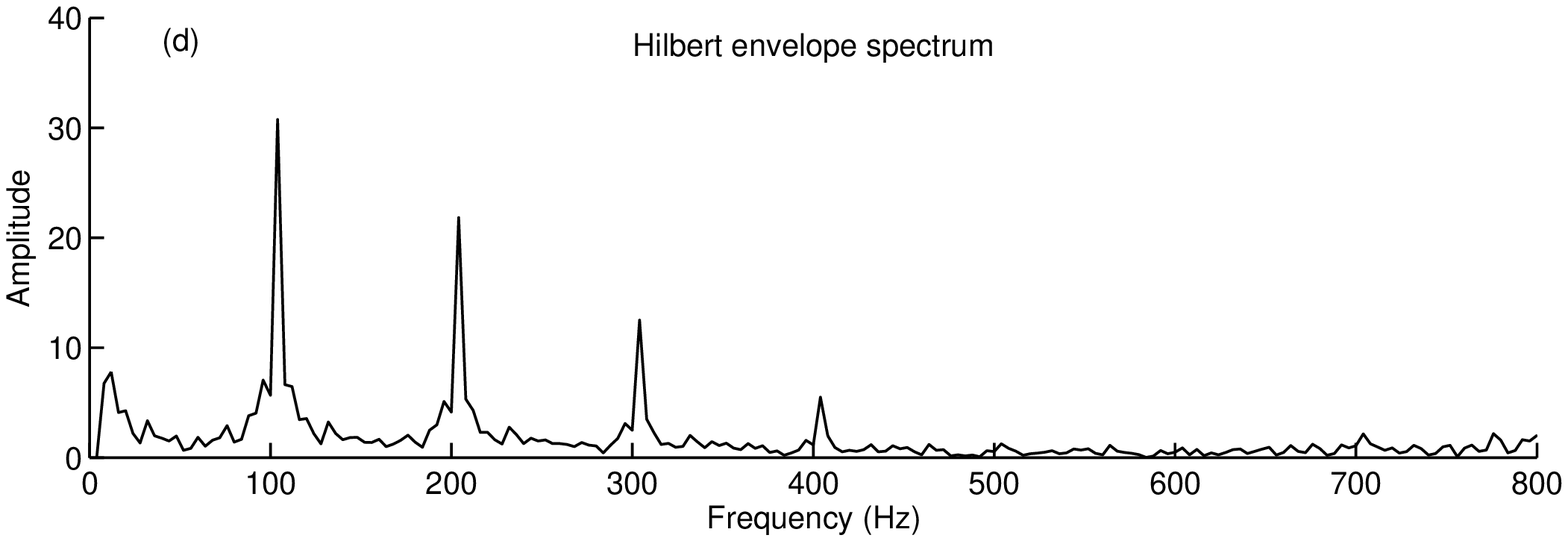}		
	\caption{Result from convex penalty function,
		(a) estimated time domain signal,
		(b) STFT coefficients,
		(c) detected oscillatory frequency components,
		(d) Hilbert envelope spectrum of estimated signal.}
	\label{fig:fs_Example_1_cn_x}
\end{figure}

\begin{figure}[t]
\centering
	\includegraphics[scale = \figurescale] {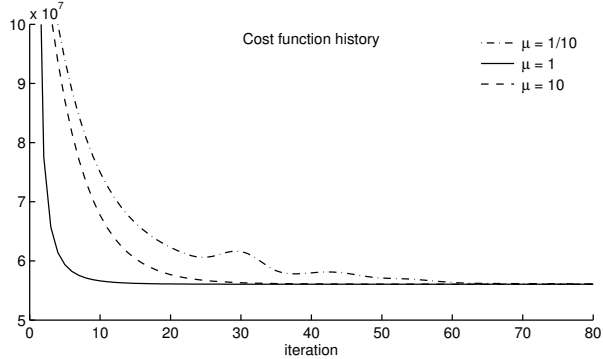}		
	\caption{Converging history of using different $\mu$'s.}
	\label{fig:fs_Example_1_mu}
\end{figure}
\begin{figure}[t]
\centering
	\includegraphics[scale = \figurescale] {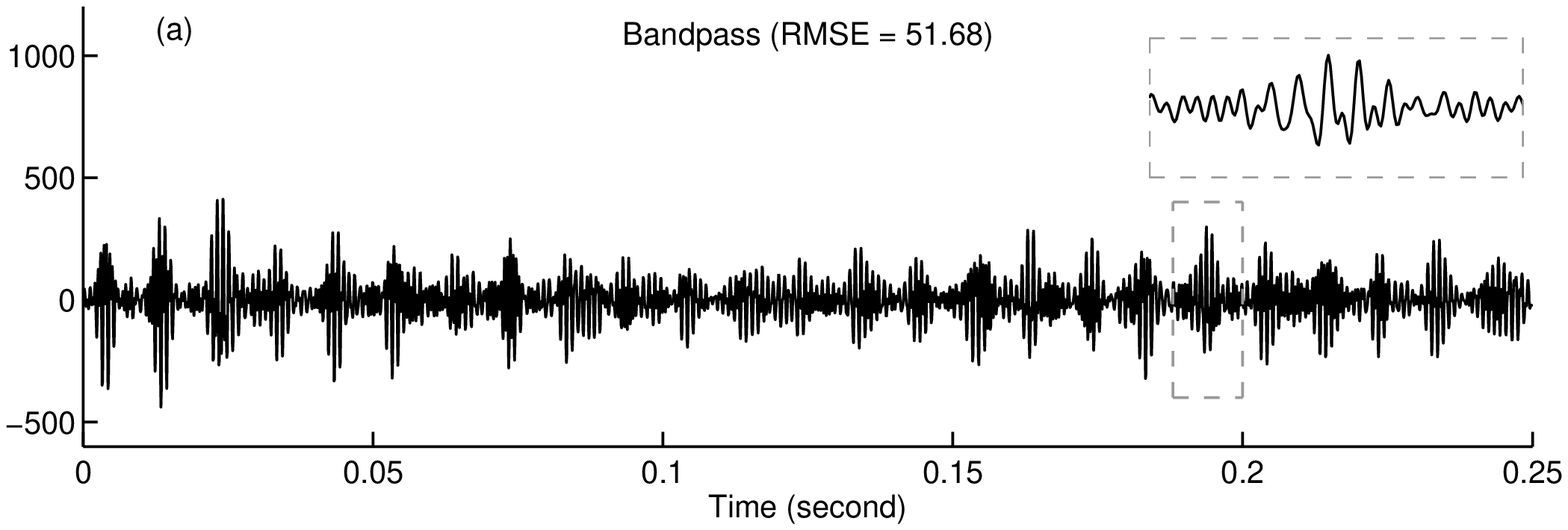} 	\\
	\includegraphics[scale = \figurescale] {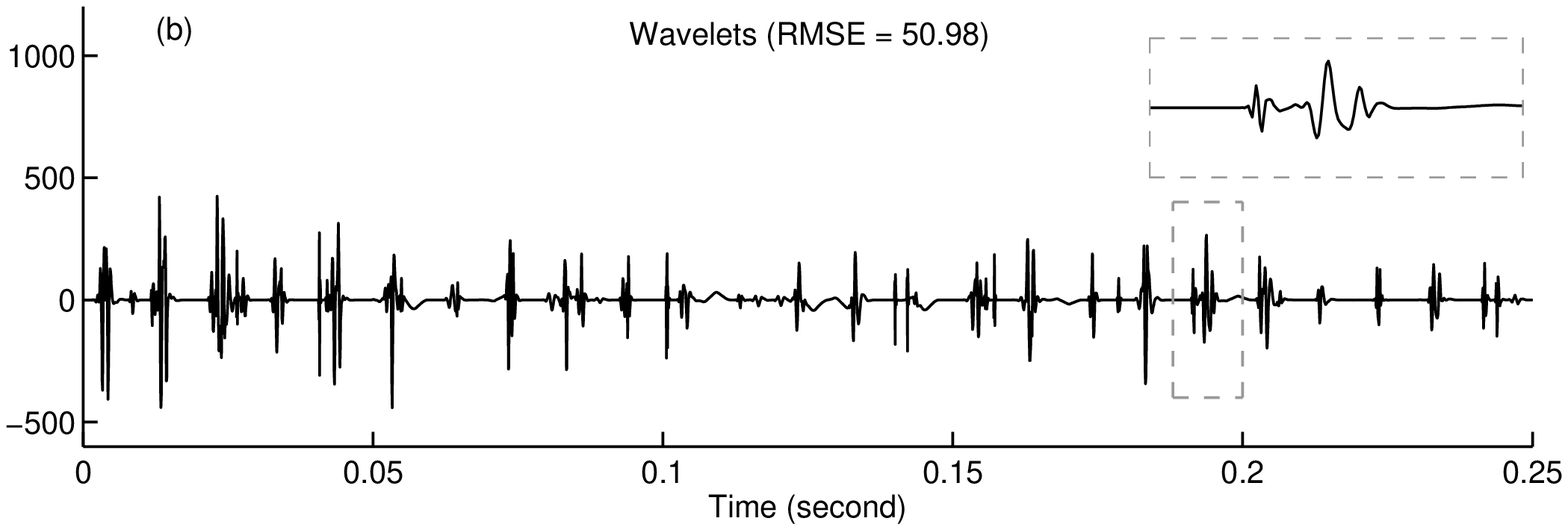}		
	\caption{Denoising results of 
		(a) bandpass filtering, 
		(b) wavelet-based denoising.}
	\label{fig:fs_Example_1_bp_x}
\end{figure}
\begin{figure}[t]
\centering
	\includegraphics[scale = \figurescale] {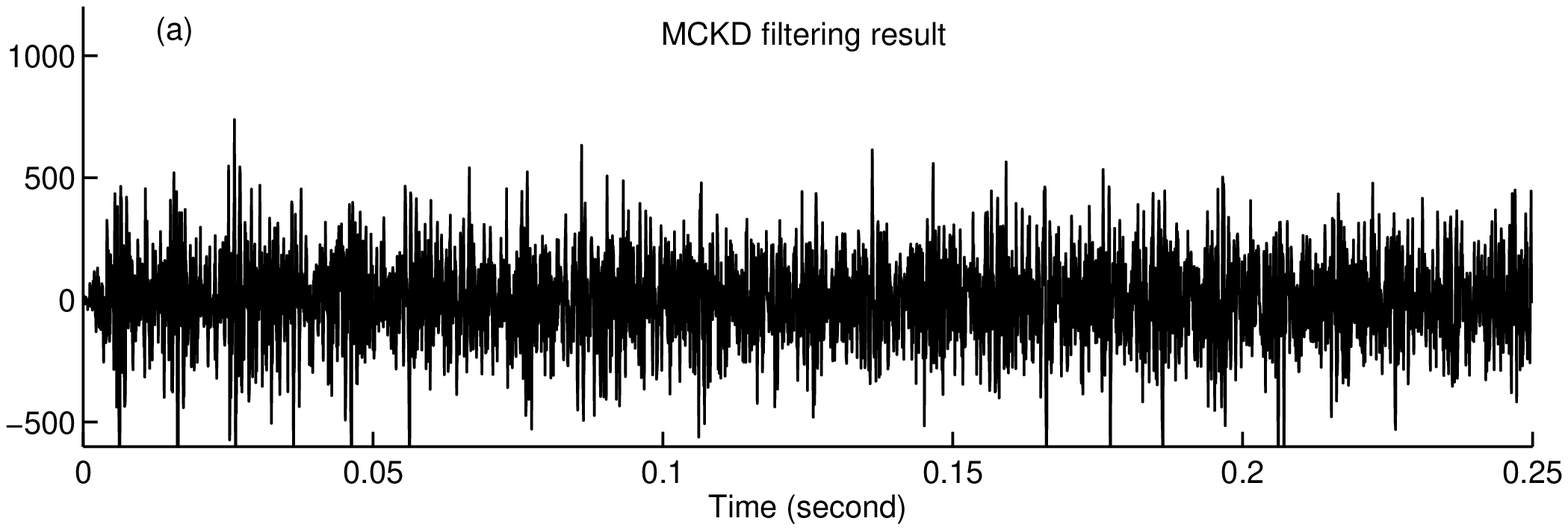}	\\			
	\includegraphics[scale = \figurescale] {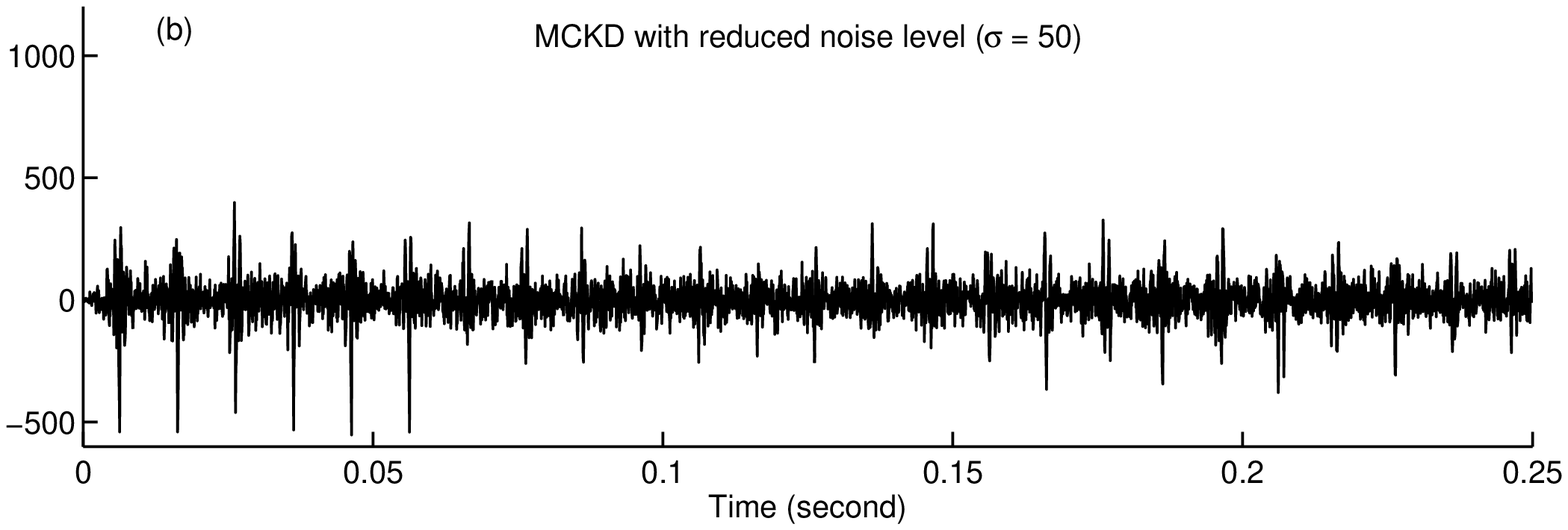}	
	\caption{Filtering result using MCKD of (a) test data, (b) test data with a reduced noise level.}
	\label{fig:fs_Example_1_mckd_x}
\end{figure}
\begin{figure}[t]
\centering
	\includegraphics[scale = \figurescale] {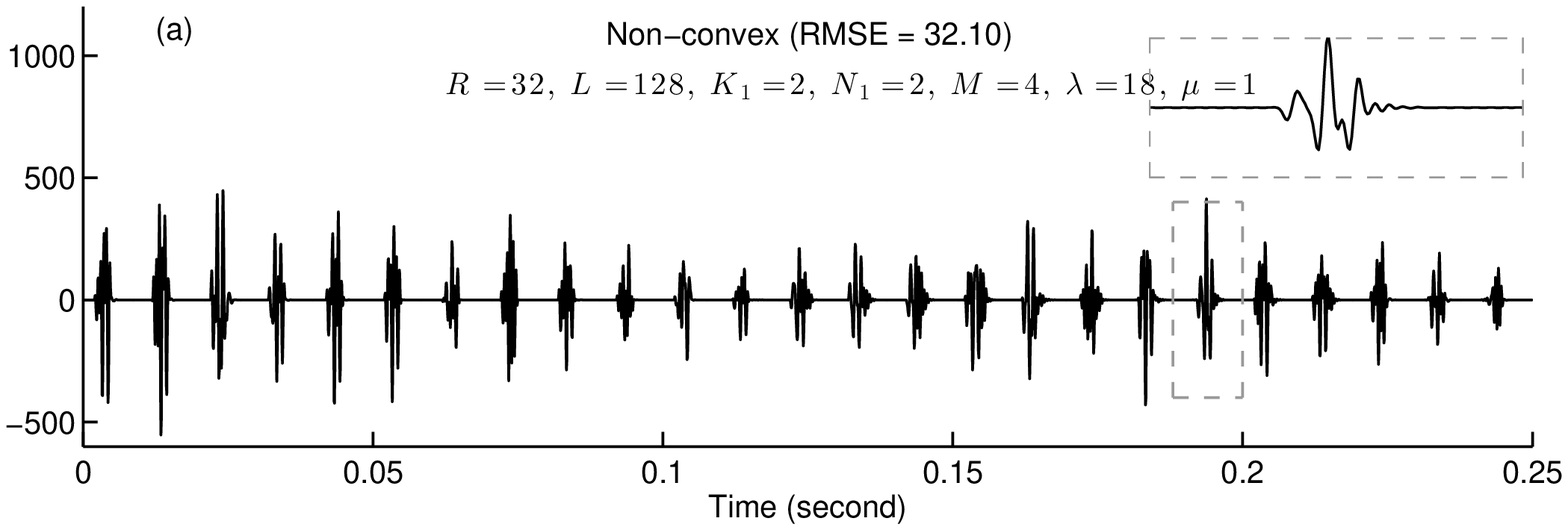}	\\	
	\includegraphics[scale = \figurescale] {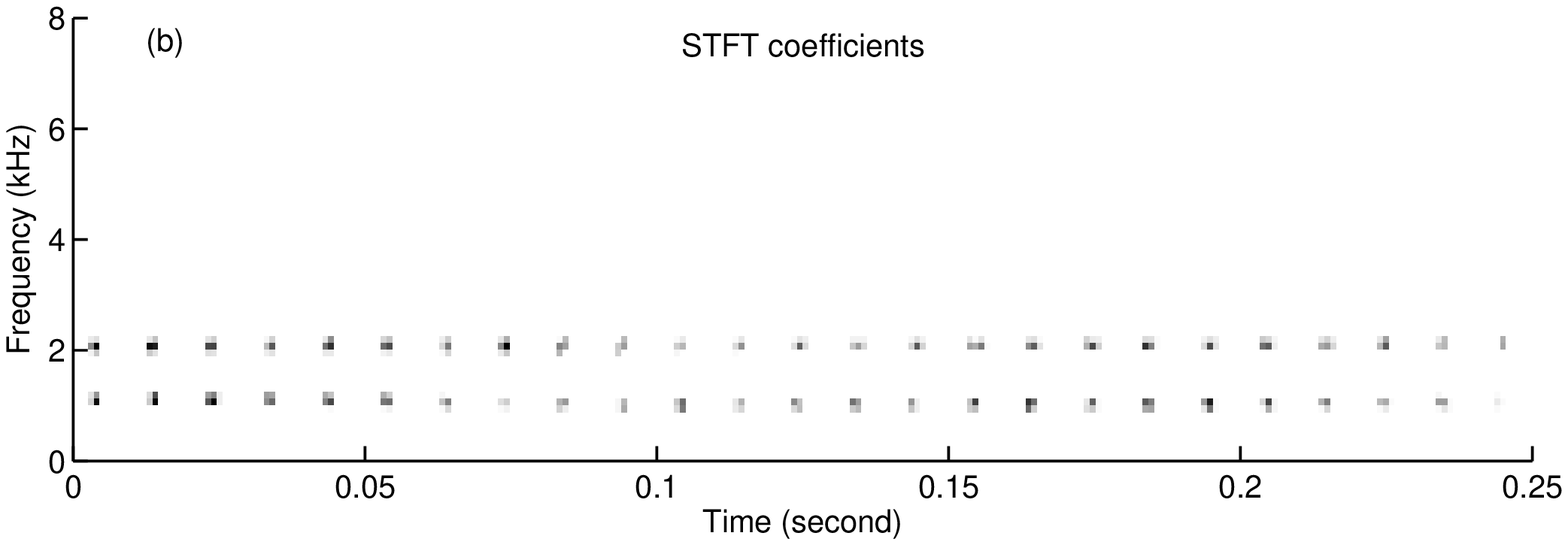}	\\	
	\includegraphics[scale = \figurescale] {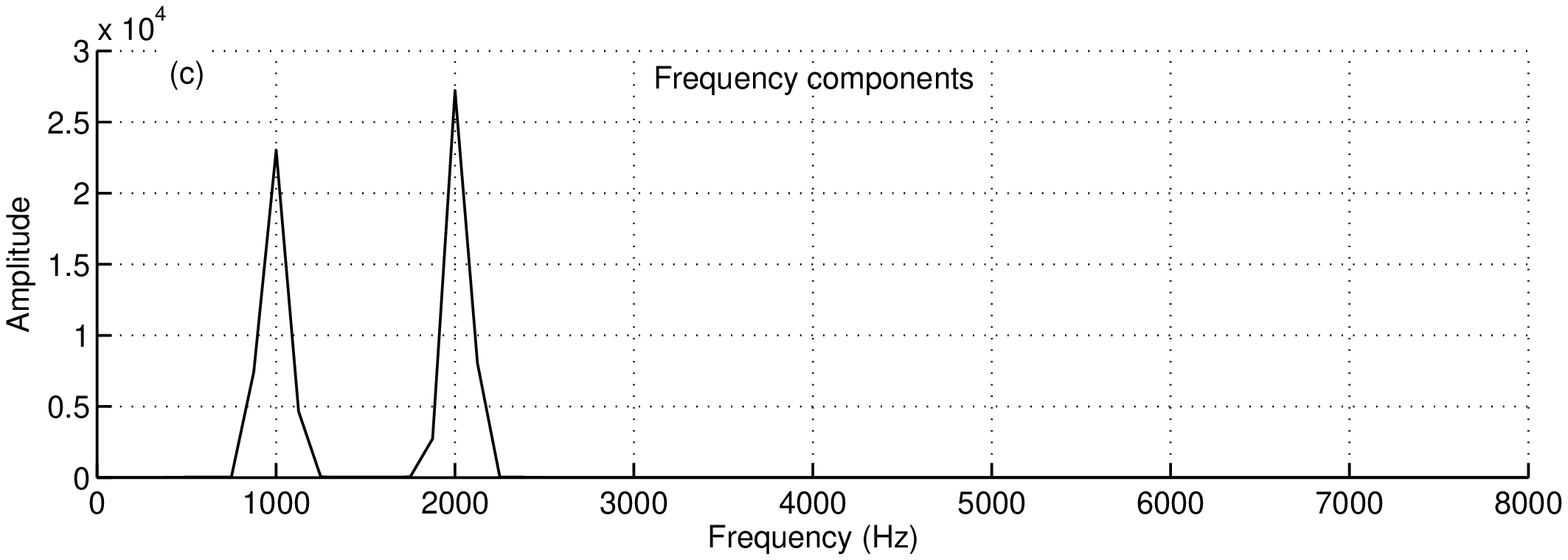}	\\	
	\includegraphics[scale = \figurescale] {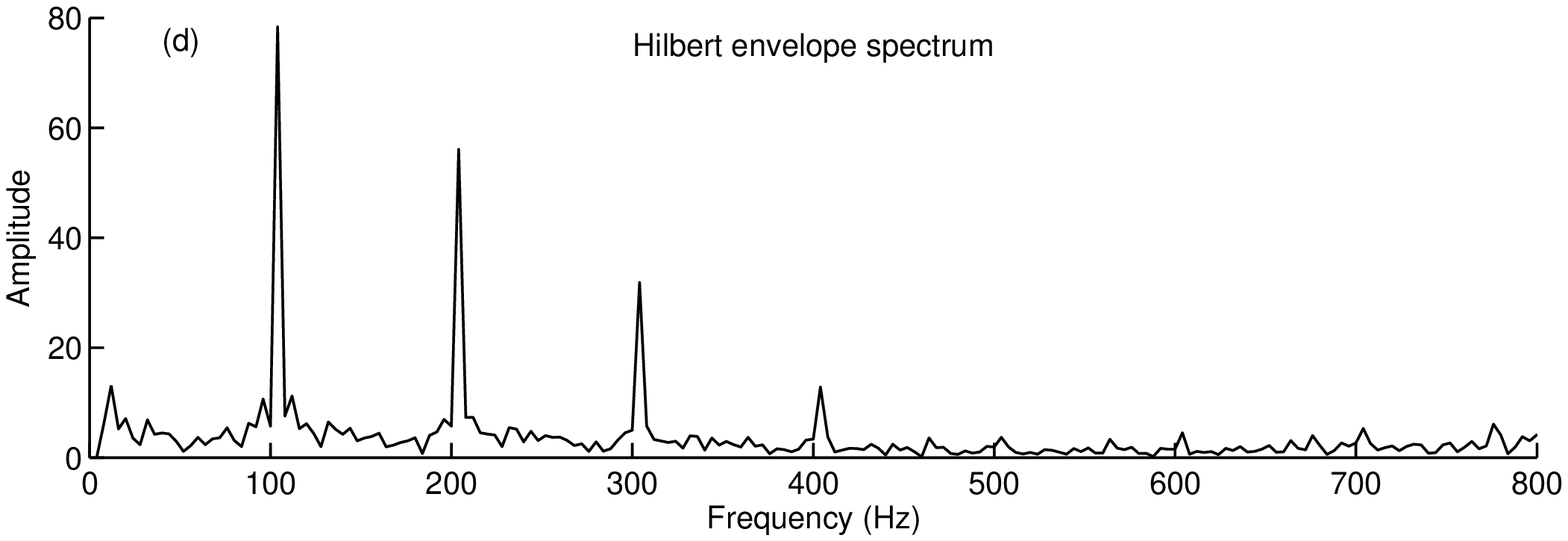}	\\		
	\caption{Enhanced result using non-convex penalty function,
		(a) estimated time domain signal,
		(b) STFT coefficients,
		(c) detected frequency components,
		(d) Hilbert envelope spectrum of estimated signal.}
	\label{fig:fs_Example_1_x}
\end{figure}

We apply the proposed method to the simulated data illustrated in \figref{fig:fs_Example_1_y}(a).
The simulated data is a 0.25 second signal with sampling rate $f_s = 16$~kHz.
We simulate the vibration features caused by fault as a transient containing two oscillation components at $1$~kHz and $2$~kHz.
Each transient lasts for about $4$ milliseconds with a random amplitude within a range,
and the two oscillatory components in one transient have independently random initial phases.
The detail of a simulated oscillatory fault is shown in \figref{fig:fs_Example_1_y}(a) (inside the dashed line box).
The clean data shown in \figref{fig:fs_Example_1_y}(a) is a sequence of the simulated faults with a period $T = 10$ milliseconds.
Note that since each transient has a random phase, in \figref{fig:fs_Example_1_y}(a),  each transient is different than the others.
We add heavy noise to the test data, and the noisy signal is shown in \figref{fig:fs_Example_1_y}(b).

In this example, we set the STFT to have a window size $ R = 32 $ with $ 50\% $ window overlapping, and the length of Fourier transform is $ L = 256 $.
We can establish the block $B$ in problem \eqnref{eqn:fs_cost}
by a condition of $b_2$ in \eqnref{eqn:fd_preoid_b} using the periodicity information,
where the length of one period in the STFT domain is
\begin{align}\label{eqn:fs_parameter}
	 \frac{ 2 T f_s }{ R } \approx N_{0} + N_{1}.
\end{align}
%
In this example, we set $K_{1} = N_{1} = 2$, and $N_{0} + N_{1} = 2 T f_s /R = 10$,
and $b_2$ as a binary weight vector is spanning $M = 4$ periods.
Hence, by \eqnref{eqn:fs_block_K}, the size of binary weight block $B$ is $ 2 \times 32 $.
When using \eqnref{eqn:fs_block} to express the block,
$B$ consists of four blocks with $ 2 \times 2$ ones and three blocks with $ 2 \times 8 $ zeros.

\figref{fig:fs_Example_1_cn_x} shows the results obtained using the proposed method with convex (smoothed $\ell_1$-norm) penalty function.
\figref{fig:fs_Example_1_cn_x}(b) is the STFT coefficients achieved by solving problem \eqnref{eqn:fs_cost},
and \figref{fig:fs_Example_1_cn_x}(a) is the estimated $\hat x$ calculated from the STFT coefficients in \figref{fig:fs_Example_1_cn_x}(b).
The proposed method recovers the oscillatory transients at accurate time and frequency.
Although the amplitude of the estimated $\hat x$ is slightly diminished, all the `fault features' (transients) are recovered at accurate locations,
and in \figref{fig:fs_Example_1_cn_x}(b), a grid of sparse blocks is distributed at the correct frequencies (about $1$ and $2$ kHz).
In this example, we set $\lam = 18$
(the method to select the parameters will be discussed in the following section).
Note that, similar to SALSA, the proposed method requires a parameter $ \mu > 0 $ which effects the convergence rate.
We use several values of $\mu$ for this example, the comparison of convergence speed is shown in \figref{fig:fs_Example_1_mu}.
Through numerical experiments, we find that setting $\mu = 1$ gives a reasonable
convergence rate, not only in this example, but also for the applications illustrated in the following sections.
To identify the frequency components, we sum the obtained STFT spectrum over time, to obtain a frequency distribution in \figref{fig:fs_Example_1_cn_x}(c),
where two peaks indicate the true oscillatory frequencies. Meanwhile, the Hilbert envelope spectrum of the estimated signal is illustrate in
\figref{fig:fs_Example_1_cn_x}(d). 
The useful fault frequency (inter-transient period) and its harmonic components are clearly revealed in the Hilbert envelope spectrum.

For comparison, we utilize bandpass filtering.
Bandpass filtering is a conventional and effective method in fault diagnosis field to extract oscillatory features or emphasize specific frequency components,
and is used in conjunction with some other techniques \cite{fd_Jiang_2007, fd_Wang_2001, fd_Sheen_JSV_2004, fd_Cong_jsv_2015}.
For instance, in \cite{fd_Wang_2001}, the author uses bandpass filtering after removing detected harmonics,
and in \cite{fd_Sheen_JSV_2004, fd_Sheen_JSV_2007}, the author demodulates vibration signals at characteristic frequency
after using specially designed bandpass filters as pre-processing,
and in \cite{fd_Cong_jsv_2015}, bandpass filtering is used as a pre-processing for a singular value decomposition (SVD) based analysis.

In this example, we use two third-order Butterworth bandpass filters with center frequencies at $1$~kHz and $2$~kHz respectively with `forward-backward' zero-phase filtering,
and the passband is about $400$~Hz for both filters.
Because the the passbands of the two filters do not overlap, we simply add the output signals from these two filters,
and the result is shown in \figref{fig:fs_Example_1_bp_x}(a).
This result shows that linear time-invariant (LTI) filtering technique hardly recovers a zero baseline.
Moreover, although some transients with large amplitude  can be visually distinguished in \figref{fig:fs_Example_1_bp_x}(a),
it is hard to distinguish the transients with small amplitudes, e.g., at about $t = 0.08$ to $0.12$ second.
Furthermore, the result in \figref{fig:fs_Example_1_bp_x}(a) is obtained by knowing the oscillatory frequencies exactly,
but in practice, especially in the area of fault diagnosis, it is rarely possible.
In this example, even though we use the accurate frequencies as prior-knowledge for the bandpass filters,
the proposed method still attains a significantly better root-mean-square error (RMSE) value.

Wavelet-based denoising has also been used to extracted fault features \cite{yan2014wavelets, fd_Yang_tpd_2001, fd_Abbasion_mssp_2007, fd_Zhen_jsv_2008}.
In this example, we adopt a conventional wavelet-coefficient thresholding method to denoise the test signal.
More specifically, a 5-scale undecimated wavelet transform \cite{Coifman_1995, LGOBW96} with 3 vanishing moments is used,
and the result is shown in \figref{fig:fs_Example_1_bp_x}(b).
For denoising, we apply hard-thresholding and chose the threshold value by $3\sigma$-rule for each subband.
In \figref{fig:fs_Example_1_bp_x}(b), although some large amplitude transients can be recovered at correct locations,
they lose the oscillatory structure, where most recovered transients have irregular waveforms
[see the cropped example inside the dashed line box in \figref{fig:fs_Example_1_bp_x}(b)].
In the time range from about $0.10$ to $0.13$ seconds, the transients are almost completely diminished.

As an application of fault diagnosis, we apply maximum correlated Kurtosis deconvolution (MCKD) \cite{fd_McDonald_mssp_2012} to the test data\footnote{
Matlab implementation is available online: \url{http://www.mathworks.com/matlabcentral/fileexchange/31326-maximum-correlated-kurtosis-deconvolution--mckd-}}.
This method deconvolves the signal from an estimated FIR filter that maximizes the correlated Kurtosis value locally.
It requires the period of potential fault features.
To proceed with the algorithm, we set the period length to be the true period (160 samples),
and the FIR filter length to be 100, and the M-shift value to be 5 as suggested in the Matlab implementation available online.
The result of MCKD is shown in \figref{fig:fs_Example_1_mckd_x}(a),
where we found that the algorithm is not able to give a deconvolution result that properly indicates most of the fault features at the noise level of our test signal.
When comparing to the noisy data in  \figref{fig:fs_Example_1_y}(b),
we can see the algorithm mistakenly promotes the impulsive features in noise.
When we reduce the noise level, the algorithm is capable to properly indicate periodic fault features as promoted impulses after deconvolution.
In \figref{fig:fs_Example_1_mckd_x}(b), we show an example of applying MCKD to the data in  \figref{fig:fs_Example_1_y}(b),
but with a reduced noise level from $\sigma = 150$ to $50$.

\subsection{Enhancement by non-convex penalty}

Non-convex penalty functions can be utilized to further enhance the effectiveness of our proposed formulation.
When the objective function \eqnref{eqn:fs_cost} has a non-convex regularizer, the problem is generally non-convex,
and when minimizing a non-convex function, the algorithm may get trapped in a local minima.
To reduce this problem, we use the solution from the convex formulation to initialize the algorithm.
Through numerical experiments, we suggest a range of values for parameter $a$, which controls the `non-convexity' of the penalty functions in \tabref{tab:etea_penalty},
\begin{align}\label{eqn:fs_suggest_a}
	0 \le a \le \frac{1}{\lam K}, \quad K = \sum_{k_1,k_2} [B]_{k_1,k_2},
\end{align}
where $K$ is also the number of $1$'s when we restrict $B$ to be a binary block as \eqnref{eqn:fs_block}.
This range does not guarantee the problem is convex,
but in practice, using the non-convex penalty regularizers in this range significantly enhances the result and gives a stable convergence behavior.

To further reduce the problem of local minima, we can gradually increase the parameter $a$ to its maximum value.
For instance, when the upper bound of $a$ is $0.25$ in \eqnref{eqn:fs_suggest_a},
we can run the algorithm (in \tabref{alg:main}) starting with $ a = 0 $ obtaining a result with convex penalty function,
and then run the algorithm with non-convex penalty function five more times initialized by previous iteration, gradually increasing $a$ by $0.05$ each time.
This simultaneously helps to restrict the non-convex solution close to the convex solution and promote stronger sparsity.

In this example, we use the arctangent function (`atan' in \tabref{tab:etea_penalty}),
and the result is shown in \figref{fig:fs_Example_1_x}(a).
There is a significant improvement reflected by the RMSE value compared to the convex penalty function in \figref{fig:fs_Example_1_cn_x}.
We also show the absolute value of STFT coefficients in \figref{fig:fs_Example_1_x}(b),
and the frequency component indication in \figref{fig:fs_Example_1_x}(c),
where significant enhancement of sparsity can be seen.
Meanwhile, the Hilbert envelope spectrum of estimated signals is illustrate in \figref{fig:fs_Example_1_x}(d). The useful fault frequency and its harmonic components are clearly detected in the Hilbert envelope spectrum.

\subsection{Parameter selection}

\textbf{Setting STFT parameters.}
Setting STFT parameters. To use the proposed method, we suggest choosing a relatively small window length because, 
assuming the length of the periodic transient is unknown, a short window length can preserve the resolution of the time-frequency spectrum. 
Moreover, in order to promote the sparsity effectively, the transform A should be over-complete. 
We suggest choosing the length of the FFT for each window (the value L in the examples) 4 or 8 times the window length (the value R in the examples). 
If $L$ is too large, it will affect the computation efficiency. 
In all the tests and examples, we consistently used the STFT with a squared-sine window with 50 \% overlapping.

\textbf{Setting binary weighting grid $B$.}
Presumably, other than the information of periodicity,
any other facts about the oscillatory feature is not given in the proposed method.
Consequently, as a lower limit of grouping scheme,
we can chose the grid on the time-frequency domain adhering $K_1 = N_1 = 2$,
so that a neighboring coefficients on both time and frequency domain
are considered.
Moreover, the value of $M$, which is the number spanning in $B$,
has no specific limits,
and in all tests and examples, we consistently set the value of $M$ to 4 or 8.

\textbf{Setting regularization parameter $\lam$.}
Based on signal model \eqnref{eqn:fs_model},
we assume the noise (excluding the periodic oscillatory transients) is white noise.
Consequently, when the observation $y$ is noise only as $w$,
the solution of problemn\eqnref{eqn:fs_cost} should be almost all zero.
In this case, if the variance of noise $\sigma_w^2$ is known,
a signal with only noise $ \hat w_n \sim \mathcal{N}(0,\sigma_w)$ 
can be built,
and the regularization parameter $\lam$ in \eqnref{eqn:fs_cost} can be trained by letting
$  x\opt = Ac\opt \approx \mathbf{0}$ through \eqnref{eqn:fs_cost} when $ y = \hat w $.
It is well-known that in most signal denoising methods using sparsity-based optimization
the regularization parameter is proportional to the noise level,
wherein the heavier the noise is, the larger regularization parameter should be utilized.
Through numerical experiments, we found that for problem \eqnref{eqn:fs_cost}, the optimal lambda is an approximately linear function of $\sigma_w$.
Moreover, it is certain that when there is no noise in the observation signal, $\lam$ should be 0,
therefore other than building a noise signal to select $\lam$ for
\eqnref{eqn:fs_cost},
alternatively it is only necessary to determine a factor $\eta$ and set $\lam \approx \eta \sigma_w$.
For guidance, we provide a table as a reference to select $\lam$  at
the variance of noise $\sigma = 1$,
under several typical settings of STFT parameters.

\begin{table}[h]
  \centering
    \caption{Factor $\eta$ for selecting $\lam = \eta \sigma$, when fixing $K_1 = N_1 = 2, \sigma = 1$.} \medskip
  \begin{tabular}{@{} l cccc @{}} 
    \toprule
    ~Parameters of STFT ~~~ &$\eta$ ($M=4$)   &$\eta$ ($M=8$) \\
    \midrule
    ~$R = 16, L = 64  $ 	& 0.120  	& 0.060 \\
    ~$R = 16, L = 128 $ 	& 0.085 	& 0.060 \\
    ~$R = 32, L = 128 $ 	& 0.120  	& 0.065 \\
    ~$R = 32, L = 256 $ 	& 0.090 	& 0.060 \\
    \toprule
  \end{tabular}
  \label{tab:fs_eta}
\end{table}

Note that in practice when the deviation of noise $\sigma_w$ is not unity,
it is just proportional to the case of $\sigma = 1$,
so it is enough to provide a suggestion of $\eta$ for $\lam = \eta \sigma$ when $\sigma = 1$ in \tabref{tab:fs_eta}.
Using the table, we can set $\lam$ by
\begin{align}\label{eqn:fs_lambda}
	\lam = \eta \cdot \alpha = \eta \cdot \sigma_w.
\end{align}
For instance, in the example shown in
\figref{fig:fs_Example_1_cn_x} and \figref{fig:fs_Example_1_x}, since we know the noise level is $\sigma_w = 150$,
then when the STFT has the parameters $R = 16, L = 64 $ and the binary grid has $M=4$,
$\lam$ can be selected by $\lam = 0.120 \times 150 = 18$.

Moreover, when the noise level is not known, we suggest to use a method given in \cite{wav_Donoho_93_ideal} to estimate the noise level,
which is
\begin{align}\label{eqn:fs_mad}
	\hat \sigma_w = \mathsf{MAD}(y) / 0.6745
\end{align}
which is a conventional estimator of noise level dependent on the noisy observation only,
used for wavelet-based denoising,
where $ \mathsf{MAD}$ is the median absolute deviation operator defined as
\begin{align}\label{eqn:fs_mad_2}
	\mathsf{MAD}(y)  :=  \mathsf{median} ( \abs{  y_n - \mathsf{median}(y)  }).
\end{align}

Note that all the examples in this paper use the above method to select parameters,
including the examples of synthetic data before this section
and the examples of experimental data (Example 2 and Example 3) following this section,
and all the specific parameters used in the examples are listed in the figures.

\section{Experiment and engineering validation}

In this section, we illustrate the proposed method by applying it to data recorded from two different machines.
One is from a public dataset concerning bearing faults.
The other data was collected by State Key Laboratory for Manufacturing and Systems Engineering of Xi'an Jiaotong University for fault diagnosis of gearbox \cite{chen2012fault}.

\subsection{Example 2: experimental data validation}

\begin{table}[t]
  \centering
    \caption{Defect frequencies of drive end bearing (multiple by running speed)} \medskip
  \begin{tabular}{@{} cccc @{}} 
    \bottomrule
    Inner race & Outer race & Cage train & Rolling element \\
    \midrule
    5.4152 & 3.5848 & 0.39828 & 4.7135 \\
    \toprule
  \end{tabular}
  \label{tab:fault_frequency}
\end{table}

\begin{figure}[t]
	\centering
    \includegraphics [scale = \Figurescale]{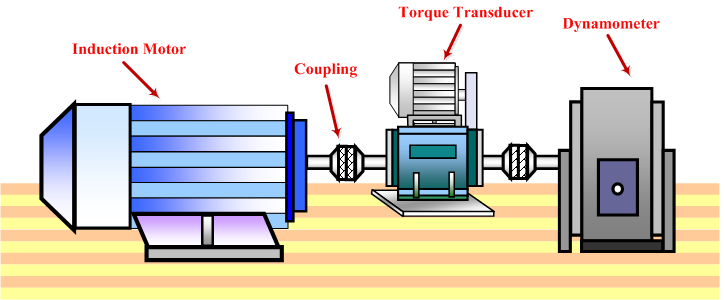}
   	\caption{Example 2: Test stand for vibration monitoring.}
	\label{fig:fs_Example_2_test_stand}
\end{figure}
\begin{figure}[t]
\centering
	\includegraphics[scale = \figurescale] {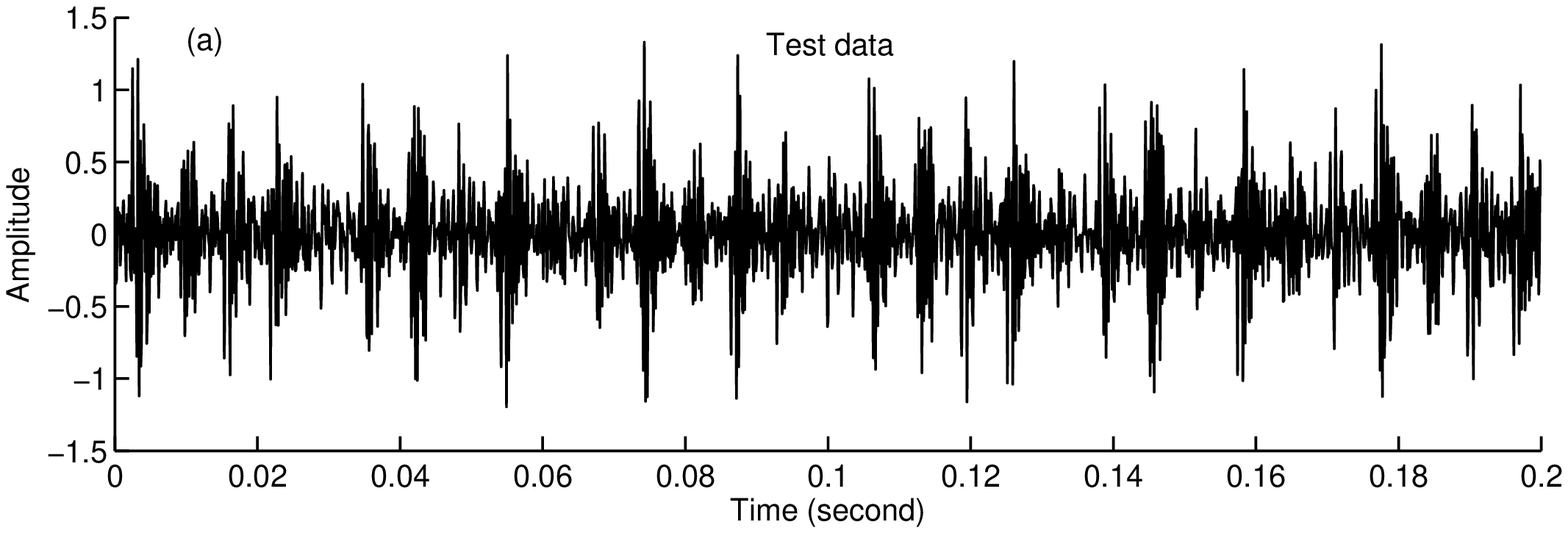}\\
	\includegraphics[scale = \figurescale] {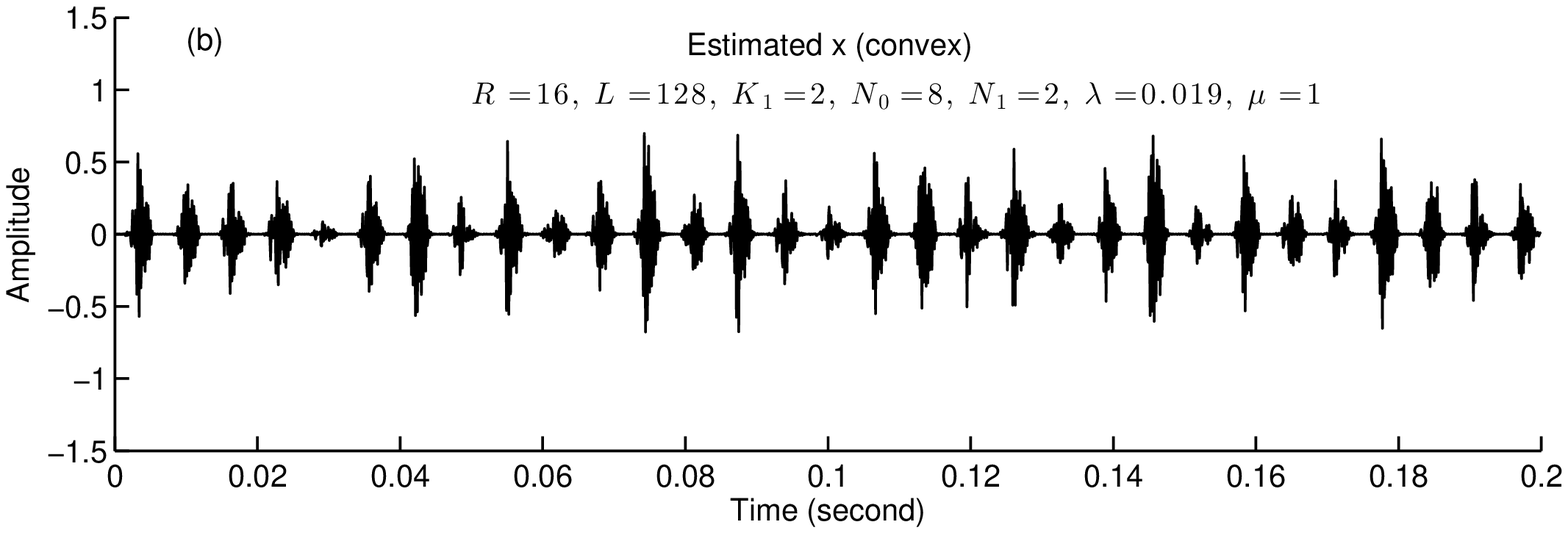}\\ \
	\includegraphics[scale = \figurescale] {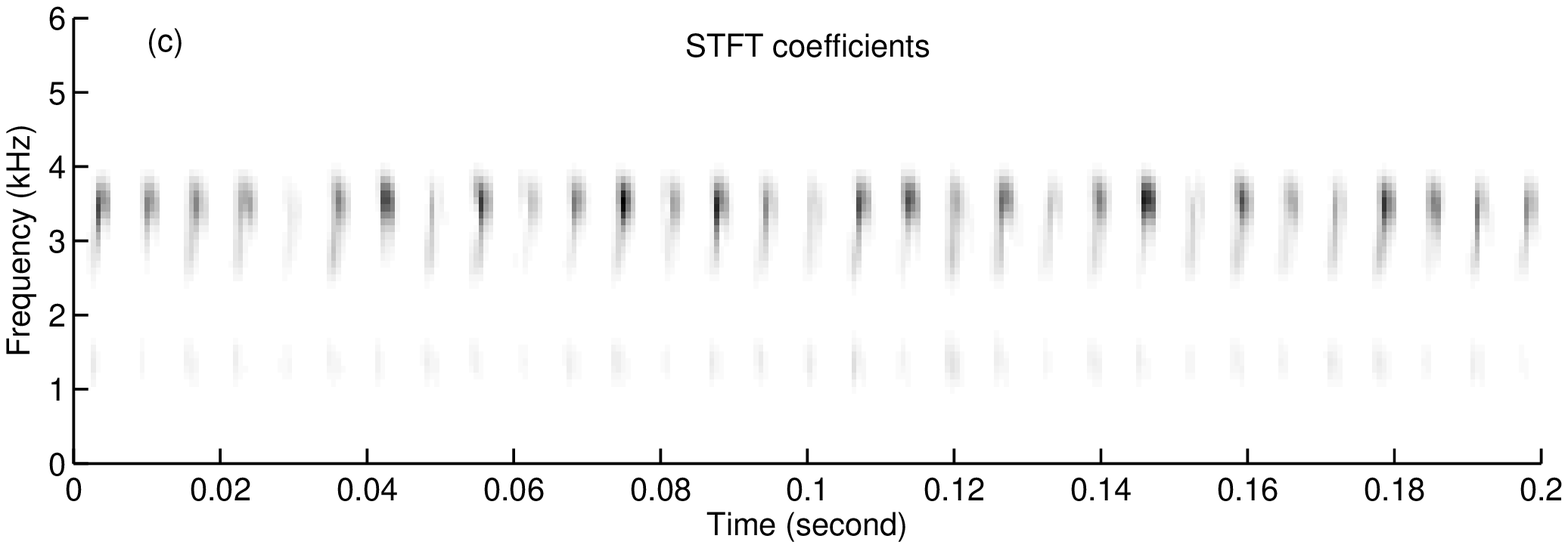}\\ \
	\includegraphics[scale = \figurescale] {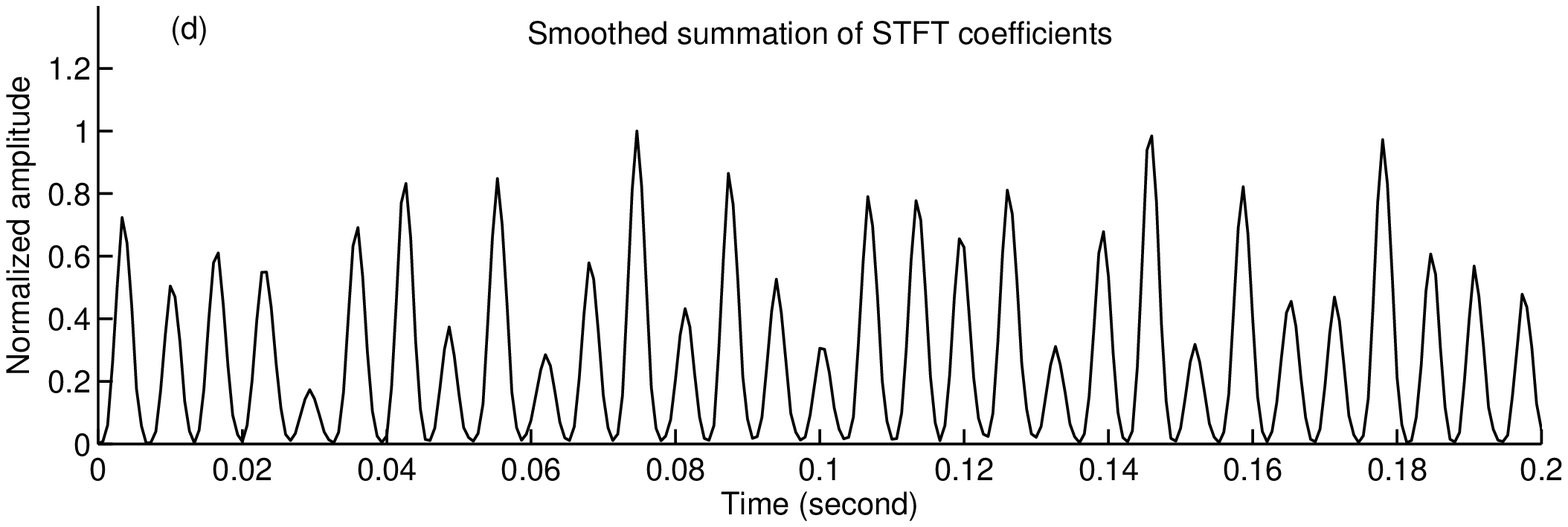}\\	
	\caption{Example 2:
		(a) test data,
		(b) estimated signal,
		(c) STFT coefficients,
		(d) smoothed profile.}
	\label{fig:fs_Example_3}
\end{figure}
\begin{figure}[t]
\centering
	\includegraphics [scale = \figurescale] {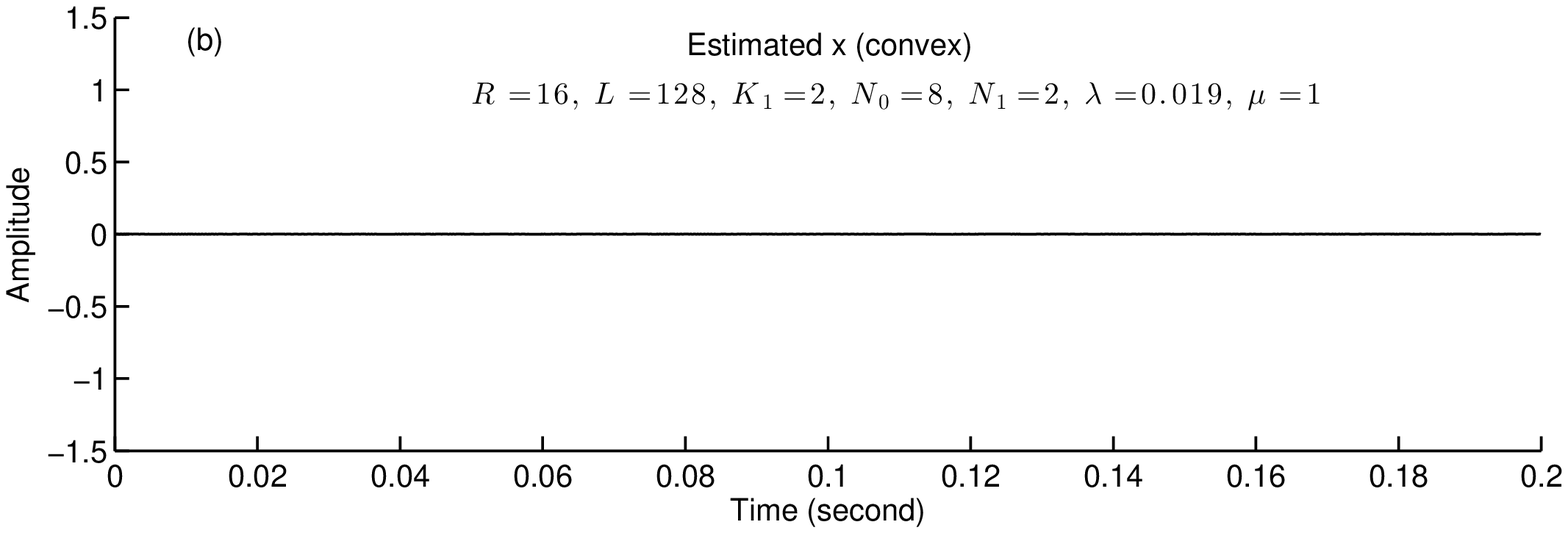}\\	
	\includegraphics [scale = \figurescale] {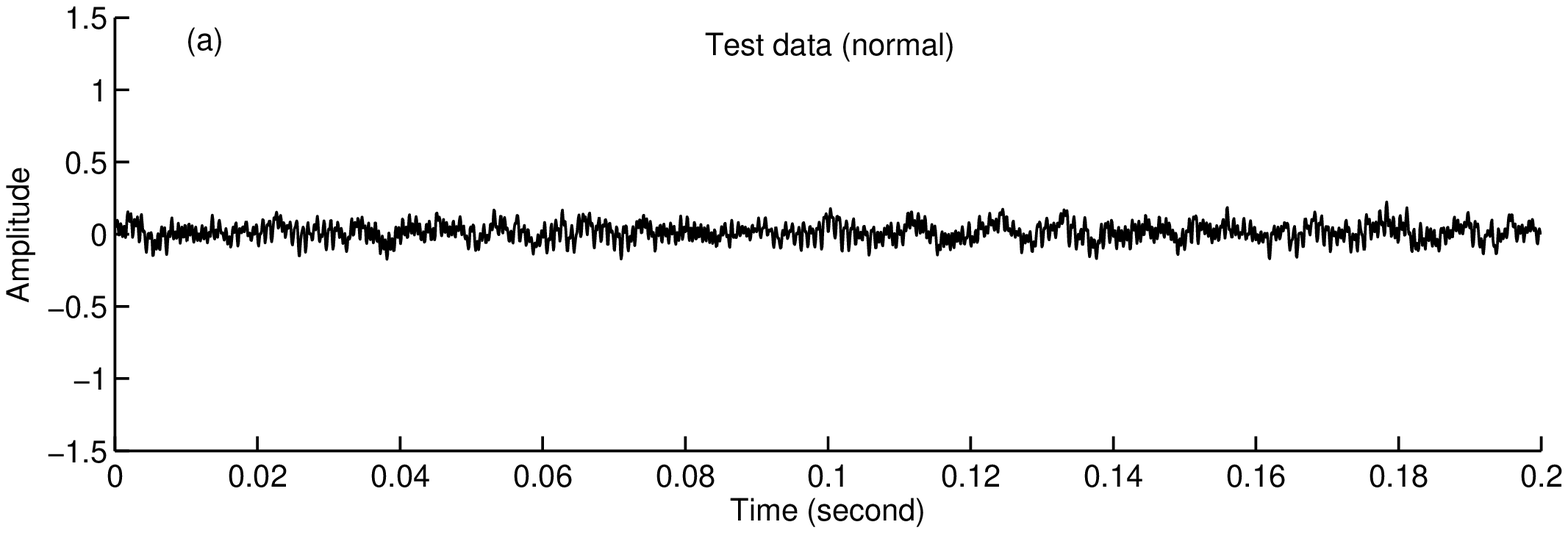}\\
	\caption{Example 2:
		(a) test data of normal setup,
		(b) estimated signal}
	\label{fig:fs_Example_3_normal}
\end{figure}
\begin{figure}[t]
\centering
	\includegraphics [scale = \figurescale] {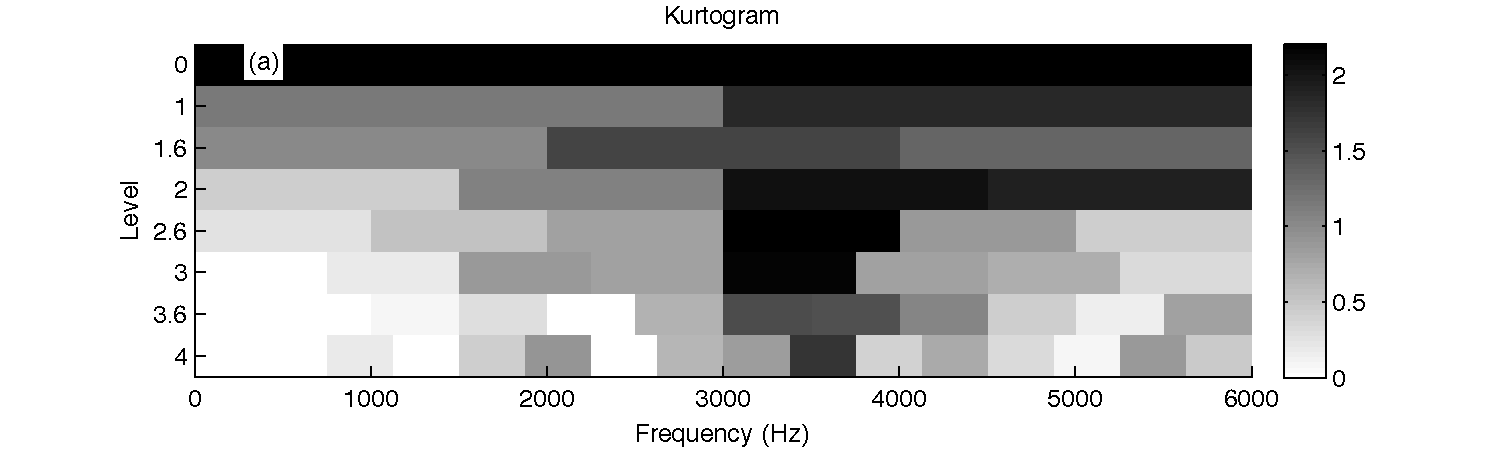}\\	
	\includegraphics [scale = \figurescale] {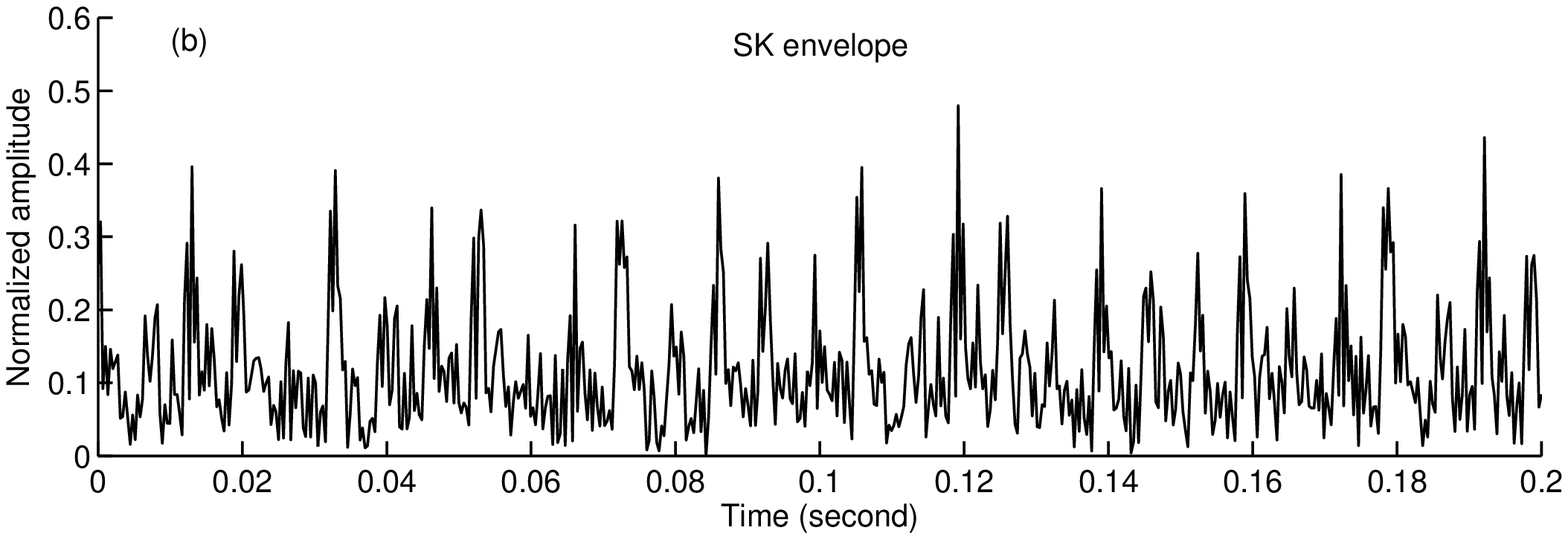}\\		
	\caption{Example 2:
		(a) Kurtogram,
		(b) envelop.}
	\label{fig:fs_Example_3_kurtogram}
\end{figure}

In this example, we utilized the proposed method to detect potential inner race fault of the drive end bearing from an experimental setup.
The experimental vibration data was acquired from the Case Western Reserve University Bearing Data Center%
\footnote{The data is available online at: \url{http://csegroups.case.edu/bearingdatacenter/pages/download-data-file}}.

As shown in \figref{fig:fs_Example_2_test_stand}, the test stand consists of a 2 hp motor (left),
a torque transducer/encoder (center), and a dynamometer (right).
The bearing installed on the drive end is a 6205-2RS-JEM SKF deep-groove ball bearing.
The potential defect frequencies of drive end bearing are listed in Table~\ref{tab:fault_frequency}, dependent on the location of the fault.
More details about the experiment setup can be found at the website \url{http://csegroups.case.edu/bearingdatacenter/pages/apparatus-procedures}.

A measured vibration signal is shown in \figref{fig:fs_Example_3}(a), with a sampling rate $f_s = 12$ kHz.
The signal was collected using accelerometers, which were attached to the housing with magnetic bases.
The current rotational speed of the motor is approximately 1730 r/min.
Thus, according to Table~\ref{tab:fault_frequency},
the fault characteristic frequency of the drive end bearing inner race is about $f_i = 1730/60 \times 5.4152 \approx 156$~Hz.
To run the proposed algorithm,
we set the STFT window length to $R = 16$ samples, and the number of the frequency bins to $128$ for each window.
For the binary block in the STFT domain, we set that each block spans $M = 4$ periods, and $K_{1} = 2$, $N_{1} = 2$,
then the block $B$ can be determined by \eqnref{eqn:fd_preoid_b}, \eqnref{eqn:fs_block} and \eqnref{eqn:fs_parameter}.
Moreover, the regularization parameter $\lam$ is selected using \tabref{tab:fs_eta} and formula  \eqnref{eqn:fs_lambda} with an estimated
$\hat \sigma_w =0.225$ calculated by \eqnref{eqn:fs_mad}.

In order to validate the reliability of the proposed algorithm, we also test it on a normal (no fault) setup under the same rotation speed.
The test signal is shown in \figref{fig:fs_Example_3_normal}(a), and using the same parameters,
there is no oscillatory feature extracted, and the output of the proposed method is almost zero [see \figref{fig:fs_Example_3_normal}(b)].

To achieve a more visualizable oscillatory feature detection signal,
we sum the absolute values of STFT coefficients $c \opt$ through frequency domain,
and then the obtained result is lowpass filtered,
\begin{align}
	s(m_2) = \textsf{LPF} \Bigg\{ \sum_{m_1} \abs{ [c\opt]_{m_1,m_2}} \Bigg\},
\end{align}
where $m_1$ denotes the frequency index and $\textsf{LPF}$ denotes a lowpass filter.
Then we can obtain a smooth time series $s \in \real^{M_2}$,
indicating the time and strength of extracted oscillatory features caused by faults [see \figref{fig:fs_Example_3}(c)].
For comparison, we utilize fast spectral kurtosis (SK) \cite{antoni2006spectral, antoni2007fast}
\footnote{The matlab implementation is available online: \url{http://www.mathworks.com/matlabcentral/fileexchange/48912-fast-kurtogram}} to analyze the same vibration data.
This method extracts the `envelope' of the optimal subband indicating potential fault features.
The Kurtogram and the envelope of the optimal subband is shown in \figref{fig:fs_Example_3_kurtogram},
where we use the optimal subband SK detected automatically at $3$~kHz and level $2.2$.
In \figref{fig:fs_Example_3_kurtogram}(b) the large spikes indicate the fault features.
The profile in \figref{fig:fs_Example_3}(d) more clearly indicates the fault features than the one in \figref{fig:fs_Example_3_kurtogram}(b).
We can apply lowpass filtering to the `noisy' envelop extracted by SK as well, but in this case,
it tends to eliminate some spikes with small amplitudes, such as the one at about $t=0.03$ second.

\begin{figure}[h]
	\centering
    \includegraphics[scale = 0.285]{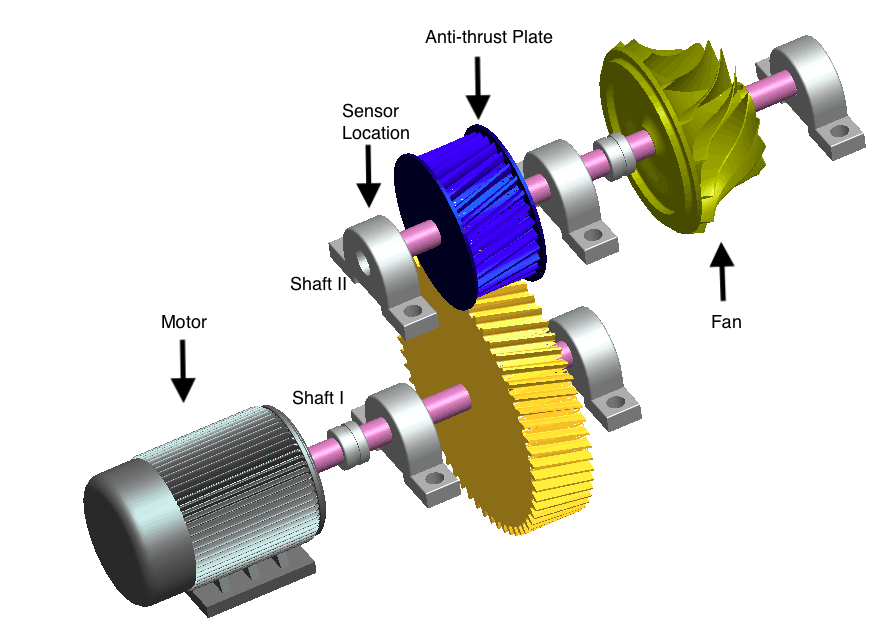}
    \includegraphics[scale = 0.275]{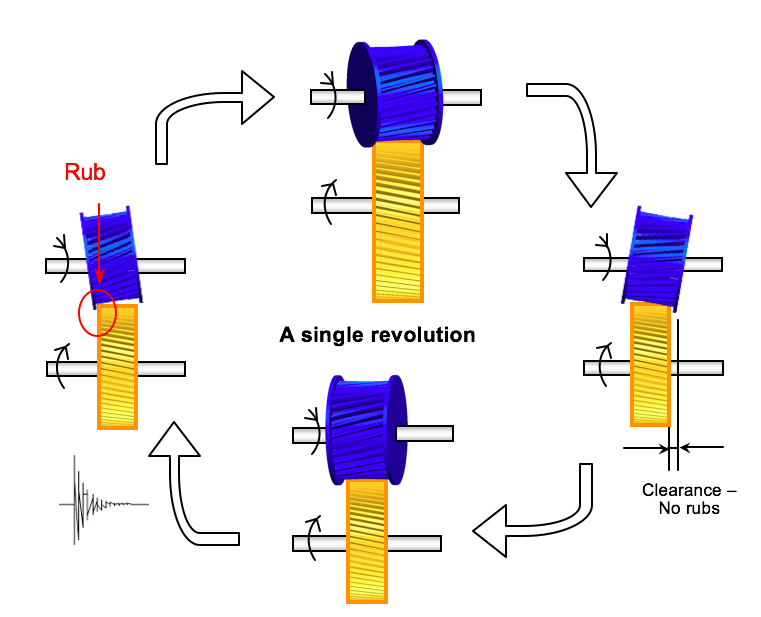}
	\caption{Example~3: Schematic of the oxygen separation and compression unit and the sensor locations (left), 
	and Illustration of rub-impact between the end face of the anti-thrust plate on the pinion and that of the bull gear (right).}
	\label{fig:oxygenunit}
\end{figure}
\begin{figure}[t]
	\centering
    \includegraphics[scale = \figurescale]{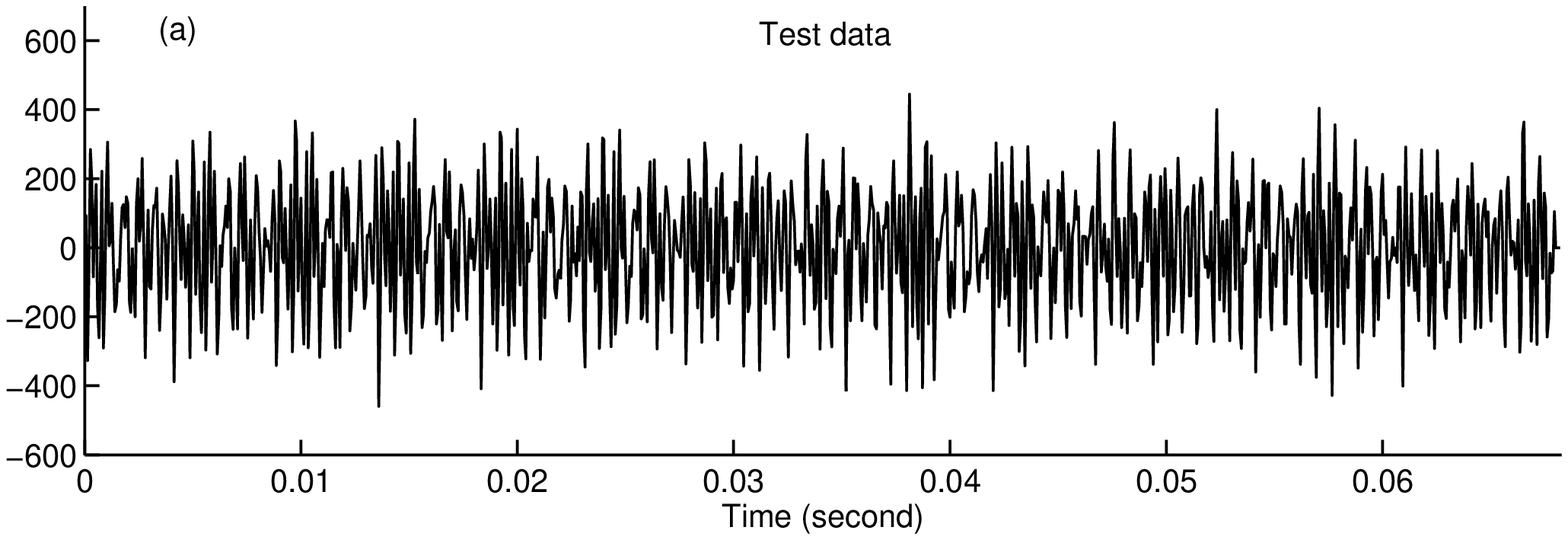} \\
    \includegraphics[scale = \figurescale]{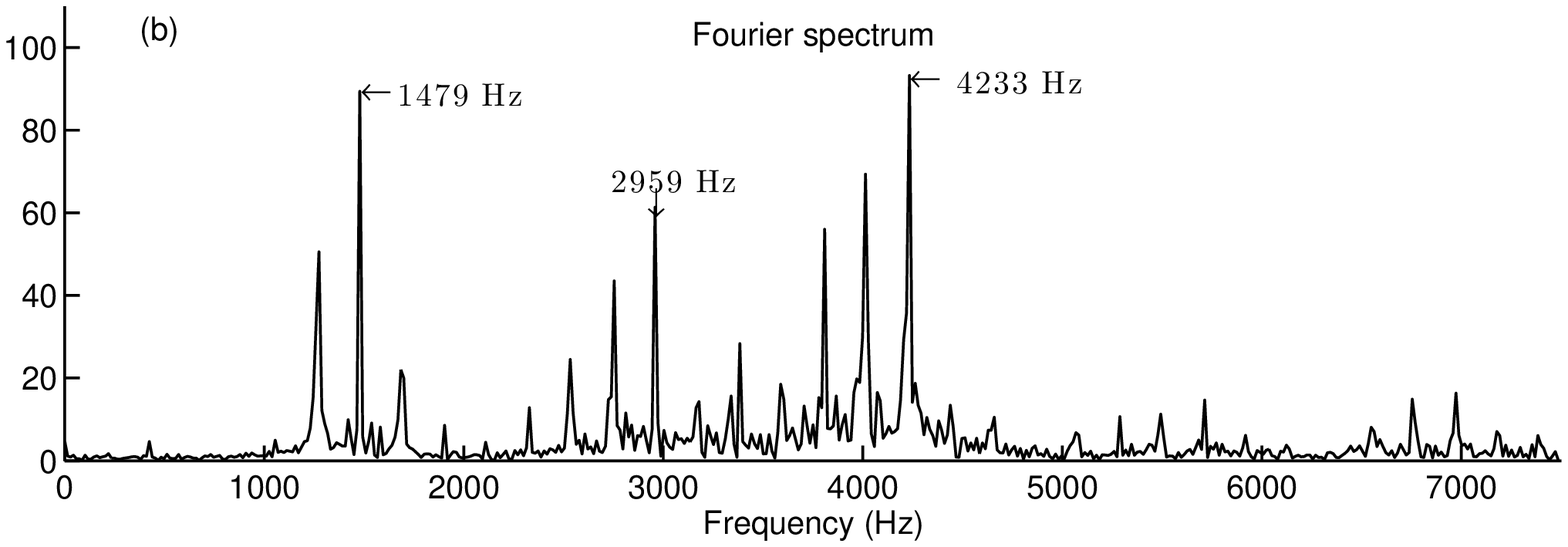} \\
	\caption{Example~3: 
				(a) measured test data.
				(b) frequency domain.}
	\label{fig:fs_Example_3_y}
\end{figure}

\subsection{Example 3: rub-impact fault detection in the gearbox of an oil refinery}

In this engineering application, we applied the proposed method to detect a rub-impact fault developed in the gearbox of an oil refinery to further verify its effectiveness.
One large oxygen separation and compression unit of this oil refinery was operating with severely abnormal sounds after an overhaul and thus suspected to have faults induced in its components.

The schematic of the oxygen separation and compression unit is illustrated in \figref{fig:oxygenunit}.
The characteristic frequencies of interest are listed in \tabref{tab:fault_frequency2}.
The sampling frequency of the acquired vibration signals is 15 kHz, and each epoch of signal is of length 1024.
One acceleration signal measured from Sensor 5 and its Fourier spectrum are displayed in \figref{fig:fs_Example_3_y}(a) and \figref{fig:fs_Example_3_y}(b), respectively.
Peak frequencies of 1479 Hz, 2947 Hz, and 4219 Hz can be observed in \figref{fig:fs_Example_3_y}(b).
In addition, the peak frequencies are all surrounded by sidebands spaced at about 213 Hz.
Note that 213 Hz exactly corresponds to the rotating frequency of shaft II,
as shown in Table~\ref{tab:fault_frequency2}. Thus, the rotating frequency of shaft II ($f_2=$ 213 Hz) can be utilized as prior knowledge of fault features.

In the practical shutdown inspection, a rub-impact fault between the end face of the anti-thrust plate on the pinion and that of the bull gear was found.
The fault was due to the inappropriate installation of the pinion \cite{chen2012fault}.
The end face of the anti-thrust plate and that of the bull gear were not strictly parallel due to non-perpendicularity between the pinion and Shaft II.
\figref{fig:oxygenunit} illustrates the rubbing fault occurred between the end face of the anti-thrust plate on the pinion and that of the bull gear.

\begin{figure}[t]
\centering
	\includegraphics[scale = \figurescale] {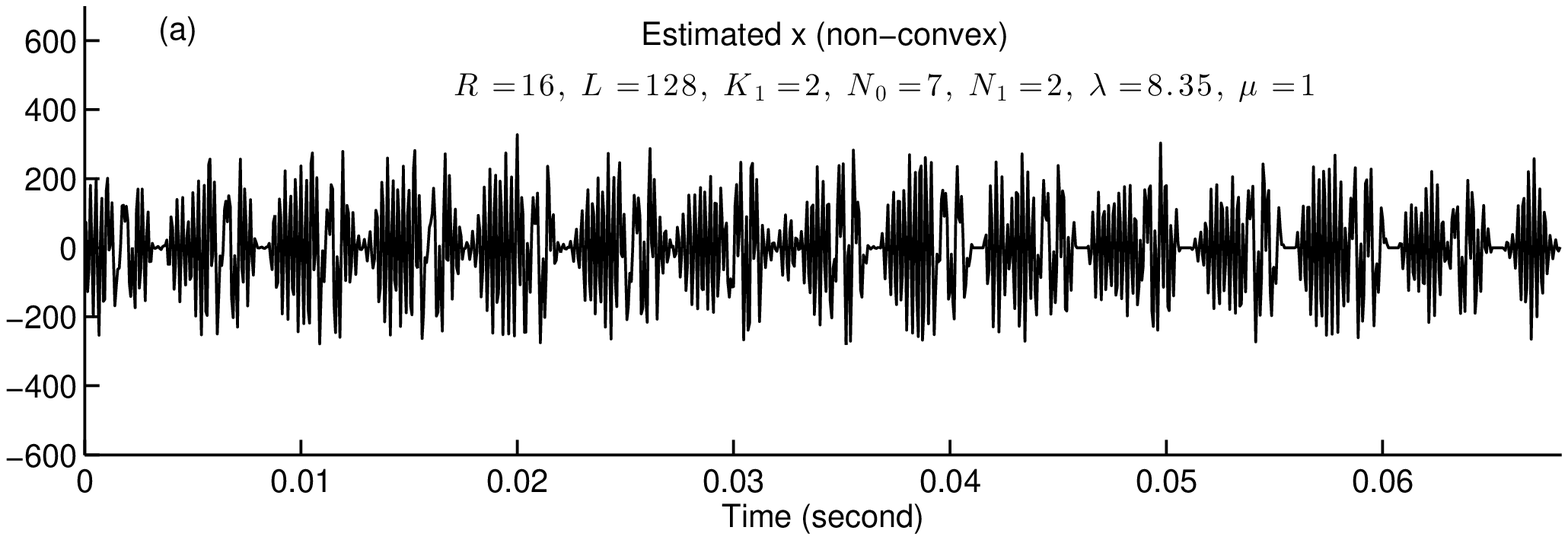} \\
	\includegraphics[scale = \figurescale] {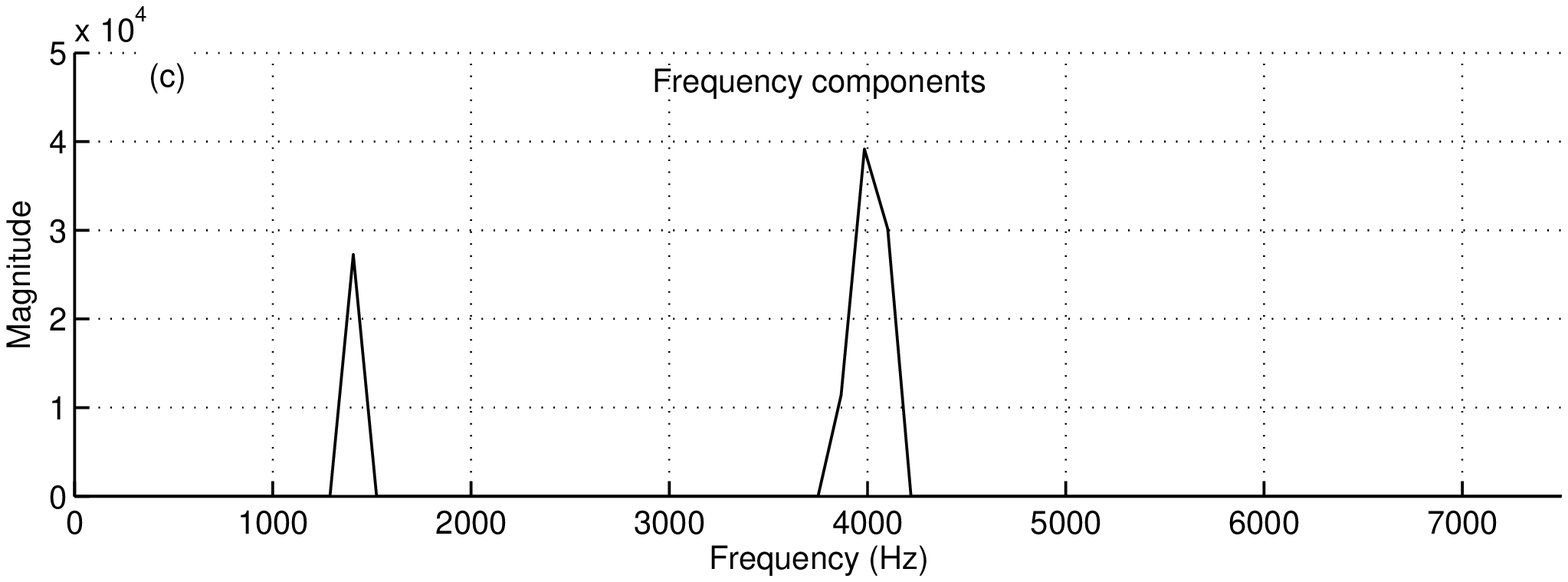} \\
	\includegraphics[scale = \figurescale] {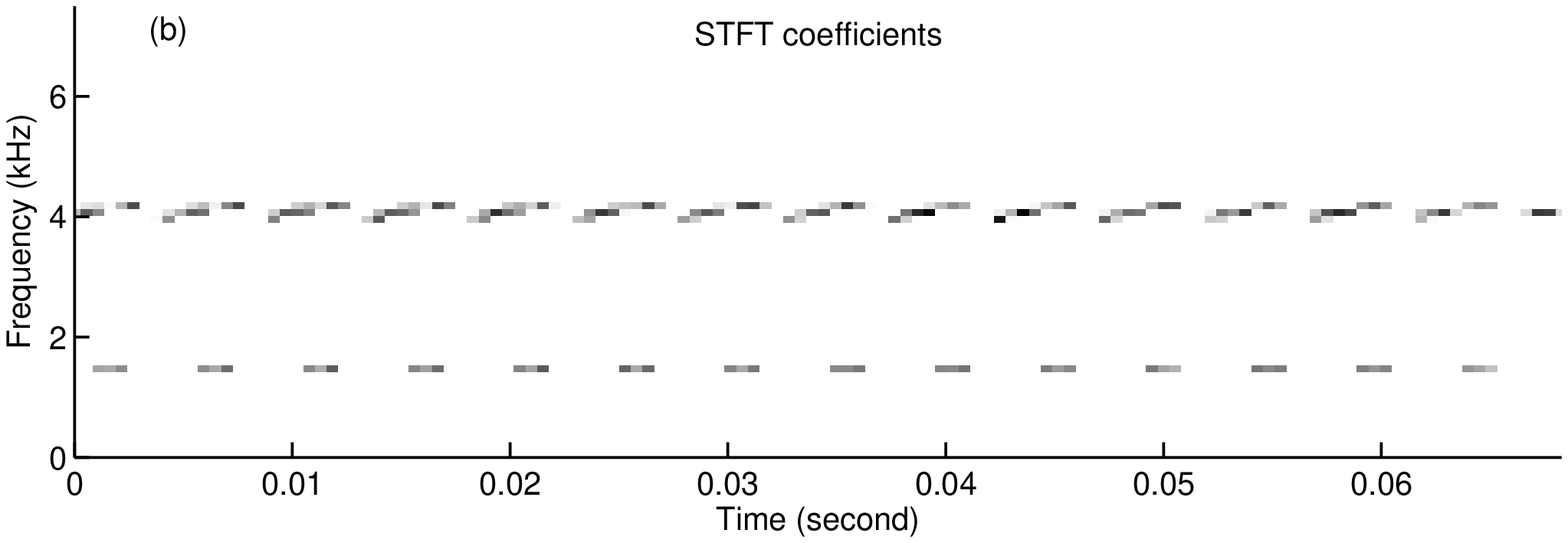} \\	
	\caption{Example~3: Result using non-convex penalty function,
		(a) estimated time domain signal,
		(b) STFT coefficients.}
	\label{fig:fs_Example_2_x}
\end{figure}

In order to capture the potential fault feature in Shaft~II,
the formula \eqnref{eqn:fs_parameter} is implemented with $T = 1/f_{2}$ to build the binary weight block \eqnref{eqn:fs_block}.
More specifically, in this example, we set $M = 8$, $K_{1} = 2$, $N_{1} = 2$, and $ N_{0} + N_{1} = 9 $,
to establish the the binary weight block $B$ with \eqnref{eqn:fd_preoid_b} and \eqnref{eqn:fs_block},
and the regularization parameter $\lam$ is determined by \eqnref{eqn:fs_lambda} with an estimated
$\hat \sigma_w =167$ calculated by \eqnref{eqn:fs_mad}.
\figref{fig:fs_Example_2_x} shows the result using proposed method with non-convex (`atan') penalty function,
where a periodic sequence of oscillatory transients is obtained with a baseline almost zero.
In the STFT domain, a sequence of periodic clusters can be observed at about $4$ kHz,
which coincides to the peak signature frequency around $4$ kHz in \figref{fig:fs_Example_3_y}(b).
The proposed method not only extracts the periodic oscillatory fault features, but also gives the oscillation frequency precisely.

\begin{figure}[t]
\centering
	\includegraphics[scale = \figurescale] {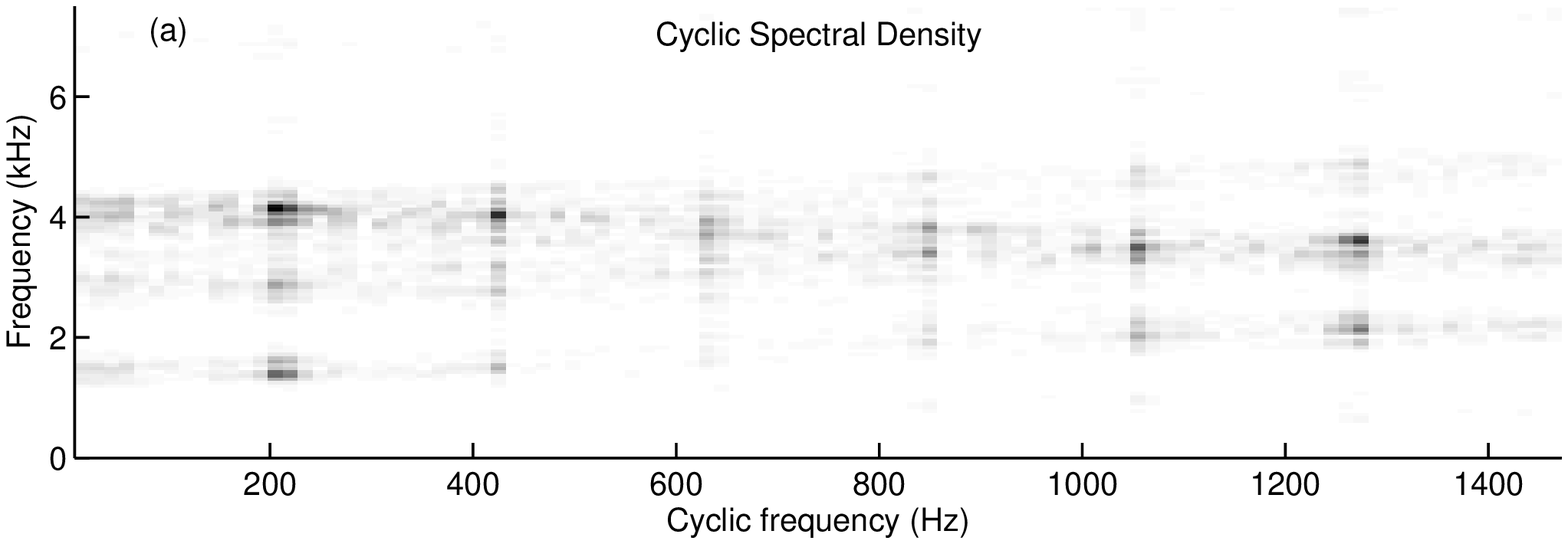} \\
	\includegraphics[scale = \figurescale] {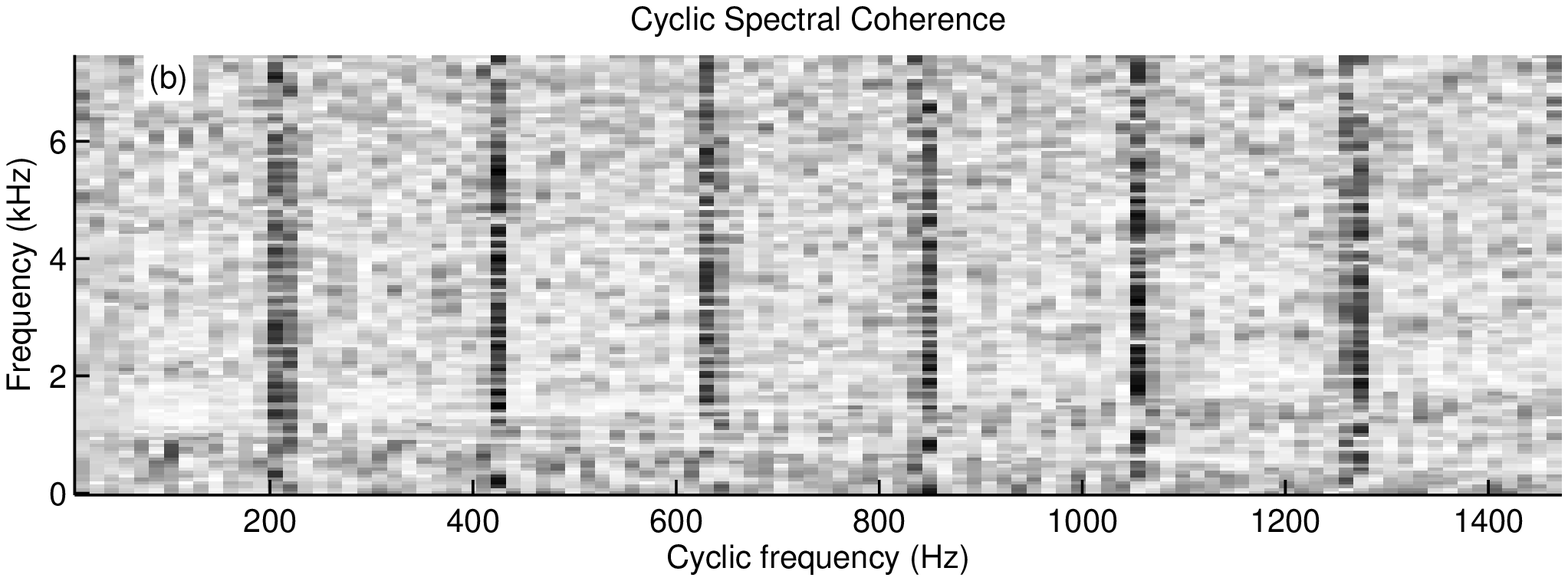} \\
	\caption{Example~3: 
		(a) Cyclic Spectral density and,
		(b) Cyclic spectral coherence of the test signal given in \figref{fig:fs_Example_3_y}.}
	\label{fig:fs_Example_2_csc}
\end{figure}

As a comparison, we tested the cyclic spectral analysis method proposed in
\cite{fd_Antoni_mssp_2007, fd_Antoni_jsv_2007, fd_Antoni_mssp_2009}%
\footnote{The matlab implementation is available online at \url{http://www.mathworks.com/matlabcentral/fileexchange/48909-cyclic-spectral-analysis}},
which is one of the `state-of-the-art' analysis method for vibration signal based on cyclostationary processes.
The results are shown in \figref{fig:fs_Example_2_csc}.
More specifically, \figref{fig:fs_Example_2_csc}(a) shows the cyclic spectral density,
where several suspicious frequency components can be observed on the density map.
In \figref{fig:fs_Example_2_csc}(a), at the location where the period (or cyclic frequency) is close to the truth (213 Hz),
two main components at about 1.5 kHz and 4 kHz respectively can be observed,
which coincides to the result given in \figref{fig:fs_Example_2_x}(c),
however there exists several other suspicious components on the map as well,
such as the one close to 3.5 kHz at cyclic frequency close to 1050 Hz and 1250 Hz.

Note that the cyclic spectral analysis method does not require a period (or cyclic frequency),
wherein a suitable range of possible cyclic frequency is needed to be defined (here we use 0 to 1500 Hz),
therefore as a more fair compassion,
we can assume that all the components observed at 213 Hz are the trustable results,
and ignore other suspicious high density points.
In this case, due to the lack of denoising procedure,
the density map are quite blurred comparing the a clear sparse indication given in \figref{fig:fs_Example_2_x}(b) and (c).
In addition, the proposed method in this work, is able to achieve a sparse time-frequency spectrum indicating not only the
frequency but also the time information, where the cyclostationary process based method does not have such feasibility.
Moreover, we found in this example,
the cyclic spectral coherence [see \figref{fig:fs_Example_2_csc}(b)] fails to indicate any frequency information.

\begin{table}[t]
  \centering
    \caption{The relevant characteristic frequencies of oxygen separation and compression unit} \medskip
  \begin{tabular}{@{} ll @{}} 
    \toprule
    Rotating speed of motor (r/m) 			& 2985 				\\
    Rotating frequency of shaft I (Hz) 		& 49.75 			\\
    Rotating frequency of shaft II (Hz) 	& \textbf{213} 		\\
    Meshing frequency of gearbox (Hz) 		& 6815.75 			\\
    Rotating frequency of fan blades (Hz) 	& 3620.86/4472.83 	\\
    \bottomrule
  \end{tabular}
  \label{tab:fault_frequency2}
\end{table}

\section{Conclusion}
This paper proposes an approach using structured sparsity in the STFT domain to extract oscillatory features caused by faults in rotating machinery.
We model the feature caused by a potential fault as a quasi-periodic sequence of oscillatory transients,
where each transient may contain multiple consistent frequency components.
The proposed approach uses smoothed (convex or non-convex) penalty functions to promote the sparsity of specially grouped STFT domain coefficients in an optimization problem.
This approach detects sparse and periodically distributed coefficients,
and the periodically oscillatory fault features can be extracted.
Using the estimated STFT coefficients, the signature frequency of the fault feature can be directly observed due to the nature of sparsity in the time-frequency domain.
To solve the proposed optimization problem efficiently,
an iterative algorithm which combines majorization-minimization and split augmented Lagrangian shrinkage algorithm has been illustrated.
When the problem is not convex, this algorithm still converges to a local optimum.
The effectiveness of the proposed approach is verified by simulation data, experimental data, and engineering data.
The results demonstrate the notable effectiveness of the proposed approach in extracting transients for detecting faults in rotating machines.

%
%

\appendix
\numberwithin{equation}{section}

\section{Proof of \theref{the:salma}}\label{app:proof}
\begin{proof}

By introducing a variable $u \in \real^{N}$ and adding an equality constraint, we can split the variable $x$,
and then problem \eqnref{eqn:bsum_problem} is equivalent to
\begin{align}\label{eqn:bsum_problem_cost}
	\{ x\opt, u\opt \} = 	& \arg \min_{x,u} \Big\{ F(x,u) =  F_0(x) + F_1(u) \Big\}\\
							& \text{subject to } x - u = 0 \nonumber.
\end{align}

This problem has an augmented Lagrangian
\begin{align}\label{eqn:bsum_problem_LA}
	L_A(x,u,\beta) = F(x,u) + \beta^{\tp}(u-x) + \frac{\mu}{2} \norm{u-x}_2^2,
\end{align}
where $\beta \in \real^{N}$, and $\mu >0$.
Moreover, we assume that $L_A$ in \eqnref{eqn:bsum_problem_LA} is finite when $x,u,\beta$ are all finite,
and the saddle point $( x\opt, u \opt, \beta\opt )$ exists and should be finite.
This implies that the saddle point $( x\opt, u \opt, \beta\opt )$ is at least a local minimizer of problem \eqnref{eqn:bsum_problem_cost}.
If function $G_1 : \real^{N} \times \real^{N} \to \real$ is continuously differentiable
and majorizes [is a upper-bound satisfying \eqnref{eqn:fs_the_condition}] of $F_1$,
then a upper-bound of $L_A$ can be found as
\begin{align}\label{eqn:bsum_problem_GA}
	G_A((x,u,\beta), (x,v,\gamma) ) = F_0(x) + G_1(u,v) + G_2((x,u,\beta), \gamma) + \frac{\mu}{2} \norm{u-x}_2^2,
\end{align}
where $G_2((x,u,\beta), \gamma)$ is a upper-bound of $\beta^{\tp} (u-x)$, defined as
\begin{align}
	G_2((x,u,\beta), \gamma) := \beta^{\tp} (u-x) + \frac{1}{2\mu} \norm{ \beta -\gamma }_2^2 \ge \beta^{\tp} (u-x), \quad \text{for all } \beta, \gamma \in \real^{N}.
\end{align}
Note that, different to the algorithms proposed in \cite{opt_Li_admm_preprint, opt_Chen_admm_2015_preprint, opt_Cui_admm_preprint},
since the augmented Lagrangian is also a function of the introduced multiplier [denoted as $\beta$ in \eqnref{eqn:bsum_problem_LA}],
BSUM can be utilized directly,
so that $G_{A}$ in \eqnref{eqn:bsum_problem_GA} satisfies the block-wise upper-bound conditions listed in \eqnref{eqn:bsum_block_condition},
when vectorizing  $\{ x,u,\beta \}$ as a single variable.
Hence, BSUM for three blocks (\tabref{alg:bsum}) can be directly applied to find the stationary point of $L_A$.
Moreover, since $L_A$ goes to extrema only when $\{ x,u,\beta \}$ goes to extrema,
BSUM will converge to a non-extremal stationary point of $L_A$, which is the saddle point (see convergence analysis in Ref.~\cite{opt_Razaviyayn_2013}).

As a consequence,  we can write the algorithm in \tabref{alg:bsum} explicitly `minimizing' $L_A$~in~\eqnref{eqn:bsum_problem_LA},
and it will converge to a saddle point of $L_A$,
\begin{subequations}\label{eqn:bsum_salma_1}
\begin{align}
	u^{(i+1)}		& = \arg \min_{u} 	G_1(u, u^{(i)}) + \beta^{\tp}(u-x) + \frac{\mu}{2} \norm{u-x}_2^2, 	\\
	x^{(i+1)}		& = \arg \min_{x}	F_0(x) + \beta^{\tp}(u-x) + \frac{\mu}{2} \norm{u-x}_2^2,  			\\
	\beta^{(i+1)}	& = \arg \min_{\beta} \beta^{\tp}(u-x) +\frac{1}{2\mu} \norm{ \beta - \beta^{(i)} }_2^2	\label{eqn:bsum_salma_1_c}		
\end{align}
\end{subequations}
where the step \eqnref{eqn:bsum_salma_1_c} can be solved explicitly as
\begin{align}
	\beta^{(i+1)} = \beta^{(i)} + \mu(x-u).
\end{align}

Similar to SALSA, we can reorganize the steps by introducing a variable $d\in \real^N$, and
then after simple variable substitutions (same to change ALM/MM version I to version II in Section~II of Ref.~\cite{Afonso_2010_TIP_SALSA}),
solving \eqnref{eqn:bsum_salma_1} for $\{ x\opt , u\opt\}$ turns out to be iteratively
computing the steps illustrated in \eqnref{eqn:bsum_salma_0}.
\end{proof}

\section{Matrix inverse lemma}\label{app:A}
The matrix inverse lemma has several forms. A common form is
\begin{align}
	\left( A+ B D^{-1} C \right)^{-1} & = A^{-1} - A^{-1}B\left( D + C A^{-1} B\right)^{-1}  CA^{-1}.
\end{align}

\section{Equivalent formulation of real values}\label{app:B}

To rewrite the algorithm into real value formulation, we consider the complex variables as an extended real value array, such that
\begin{align}\label{enq:fs_cbar}
	\bar c = \begin{bmatrix} c_{r} \\ c_{i} \end{bmatrix} = \begin{bmatrix} \Re(c) \\ \Im(c) \end{bmatrix} \in \real^{M_1 \times M_2 \times 2 },
\end{align}
and correspondingly the original variable $c$ can be recovered by $c = c_r + c_i \ii$.
Additionally, to facilitate further derivations, we also define all the other complex variables into a formulation of real values,
\begin{align}
	\bar u = \begin{bmatrix} u_{r} \\ u_{i} \end{bmatrix},
	\quad
	\bar d = \begin{bmatrix} d_{r} \\ d_{i} \end{bmatrix},
	\quad
	\bar v = \begin{bmatrix} v_{r} \\ v_{i} \end{bmatrix}.
\end{align}
Moreover, we also consider the short-time Fourier transform operator $A^{\ct}$ and its inverse operator $A$ in \eqnref{eqn:fs_cost}
into a formulation that works on real domain.
We define $ \bar A : \real^{M_1 \times M_2 \times 2} \to \real^{N}$ is a operator working equivalently to $A$ when realigning $c$ as $\bar c$ in \eqnref{enq:fs_cbar}, where
\begin{align}
	 \bar A \bar c = x,
\end{align}
and its inverse is to do the transform and then align the real and imaginary parts of the output,
\begin{align}
	 \bar A^{\tp} x = \begin{bmatrix} \Re(A^{\ct}x) \\ \Im(A^{\ct}x) \end{bmatrix},
\end{align}
Hence, the property \eqnref{eqn:fs_property} is still persevered as $ \bar A \bar A^{\tp} = I $.
As a consequence, solving the problem sequence \eqnref{eqn:fs_salma_1} is equivalent to compute the following steps,
\begin{subequations}\label{eqn:fs_salma_1_real}
\begin{align}
	u^{(i+1)} =
	& \left\{ \begin{array}{lr}
		\displaystyle
		\bar u^{(i+1)}	= ( \bar  c + \bar d ) ./ \Big[ 1 + \frac{\lam \bar r( \bar u^{(i)} ) }{\mu}   \Big] 	\label{eqn:fs_salma_1_real_a} \\[0.8em]
		u^{(i+1)} = u_r^{(i+1)} + \ii u_i^{(i+1)}
	\end{array}\right.
	\\[0.4em]
	c^{(i+1)} =
	& \left\{ \begin{array}{lr}
	\displaystyle
		\bar c^{(i+1)}	 = ( \bar  u - \bar d ) + \frac{1}{\mu+1} \bar  A^{\tp} \Big[y - \bar A( \bar u - \bar  d ) \Big] 	\\[0.8em]
		c^{(i+1)} = c_r^{(i+1)} + \ii c_i^{(i+1)}
	\end{array}\right.
	\\[0.4em]
	d^{(i+1)} =
	& \left\{ \begin{array}{lr}
	\displaystyle
		\bar d^{(i+1)} = \bar  d^{(i)} - ( \bar  u - \bar c ) \\[0.4em]
		d^{(i+1)} = d_r^{(i+1)} + \ii d_i^{(i+1)}	
	\end{array}\right.	
\end{align}
\end{subequations}
where $ \bar r( \bar v ) \in \real ^{ M_1 \times M_2 \times 2} $ in \eqnref{eqn:fs_salma_1_real_a} is defined as
\begin{align}\label{eqn:fs_rmnj}
	[\bar r( \bar v)]_{m_1,m_2,j}
	= \sum_{k_1 = 0} ^{K_1-1} \sum_{k_2 = 0} ^{K_2-1} \frac{ [B]_{k_1,k_2}^2 }{ \psi( p(\bar v, B, m_1-k_1, m_2-k_2)  ) },
	\quad \text{for } j = 1, 2.
\end{align}
where
\begin{align}
	 p(\bar v, B, j_1, j_2) =  \Big[ \norm{ B \odot S(v_r,j_1, j_2) }_2^2 + \norm{ B \odot S(v_i, j_1, j_2) }_2^2 \Big]^{1/2}.
\end{align}
Note that when comparing \eqnref{eqn:fs_rmn} to  \eqnref{eqn:fs_rmnj},
\begin{align}
	[\bar r( \bar v)]_{m_1,m_2,j} = [ r( v)]_{m_1,m_2}, \quad \text{for } j = 1, 2.
\end{align}

Therefore, each step of \eqnref{eqn:fs_salma_2} can be understood as computing corresponding real and imaginary parts as real values,
and then recovering the corresponding complex results afterwards.
When we denote the steps on complex domain as \eqnref{eqn:fs_salma_2}, it does not affect the independency of real and imaginary parts.
Hence, the convergence of using SALMA for this problem is ensured as each step solved on real domain with proper vectorization equivalently.

\bibliographystyle{plain}

\end{document}